\documentclass[twocolumn]{aastex7}
\usepackage{graphicx}
\usepackage{threeparttable}
\usepackage{booktabs}
\usepackage{longtable}
\usepackage{threeparttablex}
\usepackage{amsmath}
\usepackage{multirow}
\usepackage{pifont}

\newcommand{\cmark}{\ding{51}}
\newcommand{\xmark}{\ding{55}}

\begin{document}

\title{The Radio Properties of Extreme Coronal Line Emitters: Constraints on the Sub-parsec Environment}

\newcommand{\LCO}{\affiliation{Las Cumbres Observatory, 6740 Cortona Drive, Suite 102, Goleta, CA 93117-5575, USA}}
\newcommand{\UCSB}{\affiliation{Department of Physics, University of California, Santa Barbara, CA 93106-9530, USA}}
\newcommand{\KITP}{\affiliation{Kavli Institute for Theoretical Physics, University of California, Santa Barbara, CA 93106-4030, USA}}
\newcommand{\UCD}{\affiliation{Department of Physics, University of California, 1 Shields Avenue, Davis, CA 95616-5270, USA}}
\newcommand{\WIS}{\affiliation{Department of Particle Physics and Astrophysics, Weizmann Institute of Science, 76100 Rehovot, Israel}}
\newcommand{\OKC}{\affiliation{Oskar Klein Centre, Department of Astronomy, Stockholm University, Albanova University Centre, SE-106 91 Stockholm, Sweden}}
\newcommand{\OAPD}{\affiliation{INAF-Osservatorio Astronomico di Padova, Vicolo dell'Osservatorio 5, I-35122 Padova, Italy}}
\newcommand{\Caltech}{\affiliation{Cahill Center for Astronomy and Astrophysics, California Institute of Technology, Mail Code 249-17, Pasadena, CA 91125, USA}}
\newcommand{\GSFC}{\affiliation{Astrophysics Science Division, NASA Goddard Space Flight Center, Mail Code 661, Greenbelt, MD 20771, USA}}
\newcommand{\UMD}{\affiliation{Joint Space-Science Institute, University of Maryland, College Park, MD 20742, USA}}
\newcommand{\UCB}{\affiliation{Department of Astronomy, University of California, Berkeley, CA 94720-3411, USA}}
\newcommand{\TTU}{\affiliation{Department of Physics, Texas Tech University, Box 41051, Lubbock, TX 79409-1051, USA}}
\newcommand{\STScI}{\affiliation{Space Telescope Science Institute, 3700 San Martin Drive, Baltimore, MD 21218, USA}}
\newcommand{\UT}{\affiliation{Department of Astronomy, The University of Texas at Austin, 2515 Speedway, Stop C1400, Austin, TX 78712, USA}}
\newcommand{\IoA}{\affiliation{Institute of Astronomy, University of Cambridge, Madingley Road, Cambridge CB3 0HA, UK}}
\newcommand{\QUB}{\affiliation{Astrophysics Research Centre, School of Mathematics and Physics, Queen's University Belfast, Belfast BT7 1NN, UK}}
\newcommand{\IPAC}{\affiliation{Spitzer Science Center, California Institute of Technology, Pasadena, CA 91125, USA}}
\newcommand{\JPL}{\affiliation{Jet Propulsion Laboratory, California Institute of Technology, 4800 Oak Grove Dr, Pasadena, CA 91109, USA}}
\newcommand{\Southampton}{\affiliation{Department of Physics and Astronomy, University of Southampton, Southampton SO17 1BJ, UK}}
\newcommand{\LANL}{\affiliation{Space and Remote Sensing, MS B244, Los Alamos National Laboratory, Los Alamos, NM 87545, USA}}
\newcommand{\Tsinghua}{\affiliation{Physics Department and Tsinghua Center for Astrophysics, Tsinghua University, Beijing, 100084, People's Republic of China}}
\newcommand{\NAOC}{\affiliation{National Astronomical Observatory of China, Chinese Academy of Sciences, Beijing, 100012, People's Republic of China}}
\newcommand{\Itagaki}{\affiliation{Itagaki Astronomical Observatory, Yamagata 990-2492, Japan}}
\newcommand{\Einstein}{\altaffiliation{Einstein Fellow}}
\newcommand{\Hubble}{\altaffiliation{Hubble Fellow}}
\newcommand{\CfA}{\affiliation{Center for Astrophysics \textbar{} Harvard \& Smithsonian, 60 Garden Street, Cambridge, MA 02138-1516, USA}}
\newcommand{\UA}{\affiliation{Department of Astronomy and Steward Observatory, University of Arizona, 933 North Cherry Avenue, Tucson, AZ 85721-0065, USA}}
\newcommand{\MPA}{\affiliation{Max-Planck-Institut f\"ur Astrophysik, Karl-Schwarzschild-Stra\ss e 1, D-85748 Garching, Germany}}
\newcommand{\MPIfR}{\affiliation{Max-Planck-Institut f\"ur Radioastronomie, Auf dem H\"ugel 69, 53121 Bonn, Germany}}
\newcommand{\DSFP}{\altaffiliation{LSST-DA Data Science Fellow}}
\newcommand{\catalyst}{\altaffiliation{LSST-DA Catalyst Fellow}}
\newcommand{\HCO}{\affiliation{Harvard College Observatory, 60 Garden Street, Cambridge, MA 02138-1516, USA}}
\newcommand{\Carnegie}{\affiliation{Observatories of the Carnegie Institute for Science, 813 Santa Barbara Street, Pasadena, CA 91101-1232, USA}}
\newcommand{\TAU}{\affiliation{School of Physics and Astronomy, Tel Aviv University, Tel Aviv 69978, Israel}}
\newcommand{\Edinburgh}{\affiliation{Institute for Astronomy, University of Edinburgh, Royal Observatory, Blackford Hill EH9 3HJ, UK}}
\newcommand{\Birmingham}{\affiliation{Birmingham Institute for Gravitational Wave Astronomy and School of Physics and Astronomy, University of Birmingham, Birmingham B15 2TT, UK}}
\newcommand{\CIERA}{\affiliation{Center for Interdisciplinary Exploration and Research in Astrophysics and Department of Physics and Astronomy, \\Northwestern University, 1800 Sherman Ave., 8th Floor, Evanston, IL 60201, USA}}
\newcommand{\Bath}{\affiliation{Department of Physics, University of Bath, Claverton Down, Bath BA2 7AY, UK}}
\newcommand{\CTIO}{\affiliation{Cerro Tololo Inter-American Observatory, National Optical Astronomy Observatory, Casilla 603, La Serena, Chile}}
\newcommand{\Potsdam}{\affiliation{Institut f\"ur Physik und Astronomie, Universit\"at Potsdam, Haus 28, Karl-Liebknecht-Str. 24/25, D-14476 Potsdam-Golm, Germany}}
\newcommand{\INPE}{\affiliation{Instituto Nacional de Pesquisas Espaciais, Avenida dos Astronautas 1758, 12227-010, S\~ao Jos\'e dos Campos -- SP, Brazil}}
\newcommand{\UNC}{\affiliation{Department of Physics and Astronomy, University of North Carolina, 120 East Cameron Avenue, Chapel Hill, NC 27599, USA}}
\newcommand{\Ohio}{\affiliation{Astrophysical Institute, Department of Physics and Astronomy, 251B Clippinger Lab, Ohio University, Athens, OH 45701-2942, USA}}
\newcommand{\AAS}{\affiliation{American Astronomical Society, 1667 K~Street NW, Suite 800, Washington, DC 20006-1681, USA}}
\newcommand{\MMT}{\affiliation{MMT and Steward Observatories, University of Arizona, 933 North Cherry Avenue, Tucson, AZ 85721-0065, USA}}
\newcommand{\Geneva}{\affiliation{ISDC, Department of Astronomy, University of Geneva, Chemin d'\'Ecogia, 16 CH-1290 Versoix, Switzerland}}
\newcommand{\Steward}{\affiliation{Steward Observatory, University of Arizona, 933 North Cherry Avenue, Tucson, AZ 85721, USA}}
\newcommand{\Leiden}{\affiliation{Leiden Observatory, Leiden University, PO Box 9513, 2300 RA Leiden, The Netherlands}}
\newcommand{\PSU}{\affiliation{Department of Astronomy \& Astrophysics, The Pennsylvania State University, University Park, PA 16802, USA}}
\newcommand{\PSUa}{\affiliation{Department of Astronomy \& Astrophysics, The Pennsylvania State University, University Park, PA 16802, USA}}
\newcommand{\PSUb}{\affiliation{Institute for Computational \& Data Sciences, The Pennsylvania State University, University Park, PA 16802, USA}}
\newcommand{\PSUc}{\affiliation{Institute for Gravitation and the Cosmos, The Pennsylvania State University, University Park, PA 16802, USA}}
\newcommand{\IAIFI}{\affiliation{The NSF AI Institute for Artificial Intelligence and Fundamental Interactions, USA}}
\newcommand{\JHU}{\affiliation{Department of Physics and Astronomy, Johns Hopkins University, 3400 North Charles Street, Baltimore, MD 21218, USA}}
\newcommand{\Utah}{\affiliation{Department of Physics \& Astronomy, University of Utah, Salt Lake City, UT 84112, USA}}
\newcommand{\UIUC}{\affiliation{Department of Astronomy, University of Illinois, 1002 W. Green St., Urbana, IL 61801, USA}}
\newcommand{\Maryland}{\affiliation{Department of Astronomy, University of Maryland, College Park, MD 20742-2421, USA}}
\newcommand{\keck}{\affiliation{W.~M.~Keck Observatory, 65-1120 M\=amalahoa Highway, Kamuela, HI 96y43-8431, USA}}
\newcommand{\cbpf}{\affiliation{Laboratório de Inteligência Artificial, Centro Brasileiro de Pesquisas Físicas, 138 Rua Dr. Xavier Sigaud 150, CEP 22290-180, 139 Rio de Janeiro, RJ, Brazil}}
\newcommand{\UFRJ}{\affiliation{Instituto de Física, Universidade Federal do Rio de Janeiro (UFRJ), Caixa Postal 68528, 21941-972 Rio de Janeiro, Brazil}}
\newcommand{\Monash}{\affiliation{School of Physics and Astronomy, Monash University, Clayton, Victoria 3800, Australia}}
\newcommand{\UCSD}{\affiliation{Department of Astronomy \& Astrophysics, University of California, San Diego, 9500 Gilman Drive, MC 0424, La Jolla, CA 92093-0424, USA}}
\newcommand{\Northwestern}{\affiliation{Department of Physics and Astronomy, Northwestern University, Evanston, IL 60208, USA}}
\newcommand{\berkeley}{\affiliation{Department of Astronomy, University of California, Berkeley, CA 94720-3411, USA}}
\shorttitle{The Radio Properties of ECLEs}
\shortauthors{Franz \& Friends}

\correspondingauthor{Noah Franz}

\author[orcid=0000-0003-4537-3575,gname=Noah,sname=Franz]{Noah Franz}
\email[show]{nfranz@arizona.edu}
\UA

\author[orcid=0000-0002-8297-2473,gname=Kate,sname=Alexander]{Kate~D.~Alexander}
\UA
\email{kdalexander@arizona.edu}

\author[orcid=0000-0003-0528-202X,gname=Collin,sname=Christy]{Collin~T.~Christy}
\UA
\email{collinchristy@arizona.edu}

\author[orcid=0000-0003-1792-2338,gname=Tanmoy,sname=Laskar]{Tanmoy~Laskar}
\Utah
\email{tanmoy.laskar@utah.edu}

\author[orcid=0000-0002-9214-4428, gname=Stefanie, sname=Komossa]{Stefanie~Komossa}
\MPIfR
\email[]{skomossa@mpifr.de}

\author[gname=Enrico, sname=Ramirez-Ruiz, orcid=0000-0003-2558-3102]{Enrico Ramirez-Ruiz}
\affiliation{Department of Astronomy and Astrophysics, UCO/Lick Observatory, University of California, 1156 High Street, Santa Cruz, CA 95064, USA}
\email{enrico@ucolick.org}

\author[gname=Jean, sname=Somalwar, orcid=0000-0001-8426-5732]{Jean Somalwar}
\berkeley
\affiliation{Kavli Institute for Particle Astrophysics \& Cosmology, P.O. Box 2450, Stanford University, Stanford, CA 94305, USA}
\email{jsomalwar@berkeley.edu}

\author[orcid=0000-0002-9392-9681,gname=Edo,sname=Berger]{Edo Berger}
\CfA
\email{eberger@cfa.harvard.edu}

\author[0000-0002-7706-5668]{Ryan Chornock}
\berkeley
\affiliation{Berkeley Center for Multi-messenger Research on Astrophysical Transients and Outreach (Multi-RAPTOR), University of California, Berkeley, CA 94720-3411, USA}
\email{chornock@berkeley.edu}

\author[orcid=0000-0002-3137-4633,gname=Fabio,sname=De~Colle]{Fabio~De~Colle}
\affiliation{Instituto de Ciencias Nucleares, Universidad Nacional Aut{\'o}noma de
M{\'e}xico, A. P. 70-543 04510 D.F. Mexico}
\email{fabio@nucleares.unam.mx}

\author[orcid=0009-0008-5392-4190,gname=Gavin,sname=Farley]{Gavin Farley}
\Utah
\email{gavinfarley04@gmail.com}

\author[orcid=0000-0001-9570-0584,gname=Megan,sname=Newsome]{Megan Newsome}
\UT
\email{megan.newsome@austin.utexas.edu}

\author[gname=B.~Ashley, sname=VanderLey,orcid=0000-0003-1152-518X]{B.~Ashley~VanderLey}
\affiliation{National Science Foundation, 2415 Eisenhower Avenue, Alexandria, VA 22314, USA}
\email{bevinashley@gmail.com}

\begin{abstract}
    A tiny fraction ($\ll1\%$) of galaxies display luminous, high-ionization metal emission lines, which may be persistent or variable. These extreme coronal lines (ECLs) are produced when soft X-ray photons intercept dense gas ($n\gtrsim10^{6-7}~{\rm cm^{-3}}$). The high X-ray flux required implicates intense nuclear activity, likely originating from tidal disruption events (TDEs) and active galactic nuclei (AGN). As ECLs are rarely seen even within these classes, their production may also require specific environmental conditions, but the details remain unclear (e.g., the geometry and volume filling factor of the ECL-producing gas). Here, we present the radio properties of a population of $27$ low-redshift ($z<0.3$) ECL emitting galaxies (ECLEs), providing a unique and previously unexplored probe of the properties of the circumnuclear medium (CNM; $\lesssim1$~pc from the black hole) in these systems. We find that $\sim 50\%$ of ECLEs produce radio synchrotron emission with luminosity and evolution consistent with TDEs and/or AGN. Radio spectral modeling of four ECLEs reveals that the ECL-producing region is (1) clumpy with a low volume filling factor ($10^{-5}\lesssim f_{V}\lesssim10^{-2}$) and (2) likely distinct from the radio emitting region (implying, e.g., a clumpy toroidal geometry). For time-variable ECLEs, these are some of the first observational constraints on the CNM geometry in formerly quiescent galactic nuclei. The unique nature of ECLEs makes them an excellent high-energy laboratory to connect the physics of accretion, photoionization, and feedback in galactic nuclei, thus motivating continued multi-wavelength monitoring.
\end{abstract} 


\section{Introduction} 
Extreme coronal line (ECL) emitters (ECLEs) are defined to be galaxies exhibiting high-ionization metal emission lines (e.g., [\ion{Fe}{10}], [\ion{Fe}{7}]) with line fluxes $\gtrsim 0.2-0.3\times$ the flux of the [\ion{O}{3}] emission line \citep{wang_extreme_2012, hinkle_coronal_2024}. The first ECLE was discovered by \citet{komossa_discovery_2008, komossa_ntt_2009} in the Sloan Digital Sky Survey (SDSS), SDSS\,J095209.56+214313.3 (hereafter SDSS\,J0952). Systematic archival searches of SDSS spectra later revealed 15 more events, resulting in 16 total ECLEs discovered in SDSS \citep[][]{wang_extreme_2012, callow_rate_2024}. Subsequent long term follow-up observations of these 16 ECLEs unveiled two classes: (1) those that show time-variable ECLs (``variable ECLEs'' or vECLEs), and (2) those that show persistent ECLs \citep[``non-variable ECLEs'' or nvECLEs;][]{yang_long-term_2013, clark_long-term_2024, callow_rate_2024, kynoch_mapping_2026}. 

Since ECLs are photoionized and have an ionization energy in the soft X-ray regime, they require a high soft X-ray photon flux \citep[][]{komossa_ntt_2009}. The source of this energetic ionizing continuum is likely the hot accretion disks surrounding supermassive black holes \citep[SMBHs; e.g.,][]{komossa_discovery_2008}. As a result, the vECLEs are thought to be associated with transient nuclear activity, such as a tidal disruption event (TDE) in which the tidal forces from a SMBH overwhelm the internal binding energy of a star \citep[][]{hills_possible_1975, rees_tidal_1988, guillochon_hydrodynamical_2013}, while the nvECLEs are thought to be associated with sustained accretion from long-lasting active galactic nuclei (AGN) \citep[][]{komossa_discovery_2008, komossa_ntt_2009, yang_long-term_2013,clark_long-term_2024}. Recently, ECLs were discovered in 11 TDE candidate host galaxies (see \citealt{kynoch_mapping_2026} for a summary), solidifying the connection between TDEs and vECLEs initially hypothesized by \citet{komossa_discovery_2008}. Combining these with the 16 discovered in archival SDSS searches, there are 27 known optically-discovered ECLEs identified prior to 2024 \citep{wang_extreme_2012, yang_long-term_2013, short_delayed_2023, clark_long-term_2024, hinkle_coronal_2024, callow_rate_2025}. 

However, it remains unclear why only a small fraction of TDEs ($\lesssim 10\%$) and AGN ($\lesssim 1\%$) produce ECLs: What makes these galaxies special?~ Early photoionization models \citep[e.g.,][]{korista_origin_1989, ferguson_physical_1997, komossa_interpretation_1997} of AGN showed that two density regimes boost coronal lines, a low-density ($n \sim 10^{2}~{\rm cm^{-3}}$) regime and a high-density ($n \gtrsim 10^{6}~{\rm cm^{-3}}$) regime. Direct measurements of the gas density from density sensitive line ratios (e.g., [\ion{Fe}{7}]$\lambda$5158/[\ion{Fe}{7}]$\lambda$6087) of the ECL emitting gas in ECLEs has resulted in values of $n_\ell \gtrsim 10^{6–7} {\rm cm}^{-3}$ \citep[where $n_\ell$ denotes a number density derived from spectral line ratios, e.g.,][]{komossa_discovery_2008, wang_extreme_2012}. Therefore, one possibility is that ECLEs have particularly dense nuclei: ECLs may only be produced when the ionizing X-ray continuum interacts with regions of dense ($n_{\rm H} \gtrsim 10^{6-8}~{\rm cm^{-3}}$) gas \citep[][]{komossa_discovery_2008, komossa_ntt_2009, wang_extreme_2012, mummery_galaxy_2025, kynoch_mapping_2026}, and therefore require gas-rich environments \citep[][]{hinkle_coronal_2024}. However, the geometry of the ECL-producing gas and, relatedly, its volume filling factor are poorly understood. In particular, it is unclear whether the ECL-producing region is a (quasi-)spherical shield of dense gas, a torus of dense gas, or dense clouds.

Radio observations are a novel, complementary probe of the ambient circumnuclear medium (CNM; $r\lesssim1~{\rm pc}$). In both AGN and TDEs, the radio emission is thought to be produced by outflows propagating through the CNM, producing observable synchrotron emission. Radio observations of TDEs are particularly interesting, as they probe the CNM of previously quiescent SMBHs, providing important constraints on their host galaxies' star formation histories on otherwise unresolvable sub-parsec scales (e.g. \citealt{generozov_influence_2017}). Radio detections of TDEs and AGN that occur in ECLEs offer a rare opportunity to simultaneously probe the ionizing radiation field, the distribution of dense circumnuclear gas, the kinetic energy of the outflow, and the interaction between that outflow and its environment. This combination of diagnostics is unique and has the potential to connect the physics of accretion, photoionization, and feedback in galactic nuclei.

Past works have used multi-frequency radio observations of TDEs to constrain  the circumnuclear gas density and outflow properties \citep[e.g.,][]{de_colle_dynamics_2012,alexander_discovery_2016, eftekhari_radio_2018, anderson_caltechnrao_2020, cendes_radio_2021, cendes_mildly_2022, goodwin_radio_2023, goodwin_radio-emitting_2023, cendes_ubiquitous_2024, christy_peculiar_2024, hajela_eight_2025, somalwar_vlass_2025, goodwin_second_2025, christy_dichotomy_2026, goodwin_resolving_2026}. However, the radio properties of ECLEs have not been explored systematically. Rather, past works have only explored the radio properties of a subset of ECLEs. \citet{wang_asassn-18ap_2024} obtained targeted radio observations of the TDE-associated ECLE AT\,2018gn and found that the radio emission was diffuse and dominated by star formation rather than transient activity. \citet{somalwar_vlass_2025, somalwar_vlass_2025-1} serendipitously discovered ECLs in 2/6 of their radio-selected TDE sample. In that work, they find that the radio luminosity and spectra of these two events are consistent with the broader population of TDEs. \citet{clark_early_2026} check the NSF Karl G. Jansky Very Large Array (VLA) Sky Survey \citep[VLASS;][]{2020PASP..132c5001L,2023ApJS..267...37G}, the VLA Faint Images of the Radio Sky at Twenty-Centimeters \citep[FIRST;][]{1994ASPC...61..165B, 1995ApJ...450..559B} survey, and the Low-Frequency array two-metre sky survey \citep[LoTSS;][]{shimwell_lofar_2022} for ECLEs discovered with the Dark Energy Spectroscopic Instrument (DESI) and find no significant detections. These radio surveys are useful for searching for relativistic jets, but are too shallow to probe radio luminosities comparable to radio-quiet AGN or typical non-relativistic TDE outflows ($L_\nu \sim {\rm few}\times10^{38}$~erg/s) for most of their sample (i.e., for events at redshift $z\gtrsim0.1$). From these past works it is clear that {\it some} ECLEs do have radio outflows. However, it still remains unclear if the outflows appear similar to the radio outflows observed from TDEs and AGN, making the analysis of a larger sample of nearby ECLEs necessary.

To further characterize the radio emission from ECLEs, we compile public archival radio observations from the VLA of 27 optically-discovered ECLEs. We also present deeper, targeted follow-up observations with the VLA and Giant Metre-wave Radio Telescope (GMRT) of seven of the ECLEs for the first time, combined with publicly available targeted observations of three other ECLEs. We then discuss the four ECLEs with the best datasets in more detail, including modeling their spectral energy distributions (SEDs) and extracting constraints on the CNM properties.

The remaining sections are organized as follows: We describe our targeted observations and data collection in Section \ref{sec:data}; In Section \ref{sec:sed-model}, we model the radio SEDs for four events; In Section \ref{sec:discussion}, we discuss our results; finally, we conclude in Section \ref{sec:conclusion}. Throughout this work we assume the flat $\Lambda$CDM cosmology presented in \citet{planck_collaboration_planck_2020} and implemented in {\tt astropy} \citep[][]{astropy_collaboration_astropy_2013,astropy_collaboration_astropy_2018,astropy_collaboration_astropy_2022}. The code associated with this work is publicly available on Zenodo  \citep[][]{franz_radio-ecle_zenodo} and GitHub\footnote{\url{https://github.com/noahfranz13/radio-ecle}}. 

\section{Data \& Observations}\label{sec:data}

A review of the literature prior to 2026 reveals 27 optically-selected ECLEs discovered prior to 2024, which are summarized in Table \ref{tab:ecle-meta}.  We note that this selection criteria excludes one radio-selected ECLE \citep[VT\,2012;][]{somalwar_vlass_2025, somalwar_vlass_2025-1} and one X-ray selected ECLE \citep[eRASSt\,J012026;][]{baldini_new_2025}, as well as ECLEs recently identified with DESI \citep[e.g., ][]{clark_early_2026, ding_exploring_2025}. We defer an analysis of these events to future work. 

\begin{table*}
\centering
\caption{ECLE Sample}
\label{tab:ecle-meta}
\begin{tabular}{lllllrcp{5cm}}
\toprule
& {\bf Name} & {\bf RA} & {\bf Dec} & {\bf Discovery} & {\bf Redshift} & {\bf vECLE?} & {\bf References} \\
\midrule
\multirow{16}{*}[-1.2cm]{\rotatebox{90}{\bf Discovered Archivally}} & SDSS\,J1715 & 17:15:04.2893 & +56:47:15.8404 & 2000-09-01 & 0.191 & \xmark & {\citet{callow_rate_2024}} \\
& SDSS\,J2220 & 22:20:55.7312 & $-$07:53:17.8464 & 2001-10-21 & 0.149 & \cmark & {\citet{callow_rate_2024}} \\
& SDSS\,J1055 & 10:55:26.4177 & +56:37:13.1010 & 2002-04-09 & 0.074 & \xmark & {\citet{wang_extreme_2012}}, {\citet{clark_long-term_2024}} \\
& SDSS\,J1342 & 13:42:44.4150 & +05:30:56.1451 & 2002-04-09 & 0.037 & \cmark & {\citet{wang_extreme_2012}}, {\citet{clark_long-term_2024}} \\
& SDSS\,J0748 & 07:48:20.6668 & +47:12:14.2648 & 2004-02-20 & 0.062 & \cmark & {\citet{wang_extreme_2012}}, {\citet{clark_long-term_2024}}  \\
 & SDSS\,J1241 & 12:41:34.2561 & +44:26:39.2636 & 2004-02-27 & 0.042 & \cmark & {\citet{wang_extreme_2012}}, {\citet{clark_long-term_2024}} \\
 & SDSS\,J1459 & 14:59:26.0676 & +40:45:38.5508 & 2004-05-21 & 0.151 & \xmark & {\citet{callow_rate_2024}} \\
 & SDSS\,J0807 & 08:07:27.3157 & +14:05:37.0892 & 2005-11-08 & 0.074 & \xmark & {\citet{callow_rate_2024}} \\
 & SDSS\,J0952 & 09:52:09.5629 & +21:43:13.2979 & 2005-12-30 & 0.080 & \cmark & {\citet{komossa_discovery_2008}}, {\citet{komossa_ntt_2009}},{\citet{wang_extreme_2012}}, {\citet{clark_long-term_2024}} \\
 & SDSS\,J1350 & 13:50:01.4946 & +29:16:09.6460 & 2006-04-23 & 0.078 & \cmark & {\citet{wang_extreme_2012}}, {\citet{clark_long-term_2024}} \\
 & SDSS\,J1247 & 12:47:26.3719 & +07:05:25.0809 & 2006-05-21 & 0.104 & \xmark & {\citet{callow_rate_2024}} \\
 & SDSS\,J0938 & 09:38:01.6376 & +13:53:17.0423 & 2006-12-23 & 0.101 & \xmark & {\citet{wang_extreme_2012}}, {\citet{clark_long-term_2024}} \\
 & SDSS\,J1402 & 14:02:04.7560 & +29:39:46.8759 & 2007-03-19 & 0.196 & \xmark\tablenotemark{\dag} & {\citet{callow_rate_2024}} \\
 & SDSS\,J1238 & 12:38:29.5894 & +18:52:37.5554 & 2008-01-16 & 0.253 & \xmark & {\citet{callow_rate_2024}} \\ 
 & SDSS\,J1207 & 12:07:19.8102 & +24:11:55.8789 & 2008-01-19 & 0.050 & \xmark & {\citet{callow_rate_2024}} \\
 & SDSS\,J1458 & 14:58:49.7267	& +19:10:33.5109 & 2008-03-14 & 0.268 & \xmark & {\citet{callow_rate_2024}} \\ 
\\ \hline \\ 
\multirow{10}{*}[-1.8cm]{\rotatebox{90}{\bf Discovered Following a TDE}} & AT2017gge & 16:20:34.99 & +24:07:26.5 & 2017-08-03 & 0.067 & \cmark & {\citet{2017TNSTR.895....1T}}, {\citet{onori_nuclear_2022}, {\citet{wang_discovery_2022}}} \\
& AT2018gn & 01:46:42.45 & +32:30:29.30 & 2022-06-29 & 0.0675 & \cmark & {\citet{2018TNSTR..58....1K}, \citet{wang_asassn-18ap_2024}} \\
 & AT2018bcb & 22:43:42.871 & -16:59:08.49 & 2018-04-27 & 0.119 & \cmark & {\citet{2018TNSTR.555....1B}}, {\citet{neustadt_tde_2020}} \\
 & AT2018dyk & 15:33:08.015 & +44:32:08.20 & 2018-05-31 & 0.037 & \cmark & {\citet{2018TNSCR1182....1H}}, {\citet{2018TNSTR.987....1F}, {\citet{clark_at_2025}}} \\
 & AT2019qiz & 04:46:37.880 & -10:13:34.90 & 2019-09-19 & 0.015 & \cmark & {\citet{2019TNSCR1921....1S}}, {\citet{2019TNSTR1857....1F}, {\citet{short_delayed_2023}}} \\
 & AT2020vdq & 10:08:53.50 & +42:42:00.40 & 2020-10-04 & 0.045 & \cmark & {\citet{2020TNSTR3060....1N}}, {\citet{somalwar_first_2025}} \\
 & AT2021dms & 03:21:24.069 & -11:08:45.71 & 2021-02-21 & 0.031 & \cmark & {\citet{2021TNSTR.542....1F}}, {\citet{hinkle_coronal_2024}} \\
 & AT2021qth & 20:11:38.914 & -21:09:36.83 & 2021-06-17 & 0.081 & \cmark & {\citet{2021TNSTR2141....1F}}, {\citet{2023ApJ...955L...6Y}} \\
 & AT2021acak & 10:34:47.99 & +15:29:22.42 & 2021-10-22 & 0.136 & \cmark & {\citet{2021TNSTR3613....1T}}, {\citet{li_at2021acak_2023}} \\
 & AT2022fpx & 15:31:03.698 & +53:24:19.13 & 2022-03-31 & 0.073 & \cmark & {\citet{2022TNSCR1771....1P}}, {\citet{2022TNSTR.836....1T}, {\citet{lin_insights_2025}}} \\
 & AT2022upj & 00:23:56.86 & -14:25:23.4 & 2022-08-31 & 0.054 & \cmark & {\citet{2022TNSTR2690....1F}}, {\citet{newsome_mapping_2024}}, {\citet{callow_rate_2024}} \\
\bottomrule
\end{tabular}
\tablenotetext{\dag}{\citet{callow_rate_2025} note that while the ECLs are non-transient there is a transient component to the infrared light curve.}
\end{table*}

\begin{figure*}
    \centering
    \includegraphics[width=\linewidth]{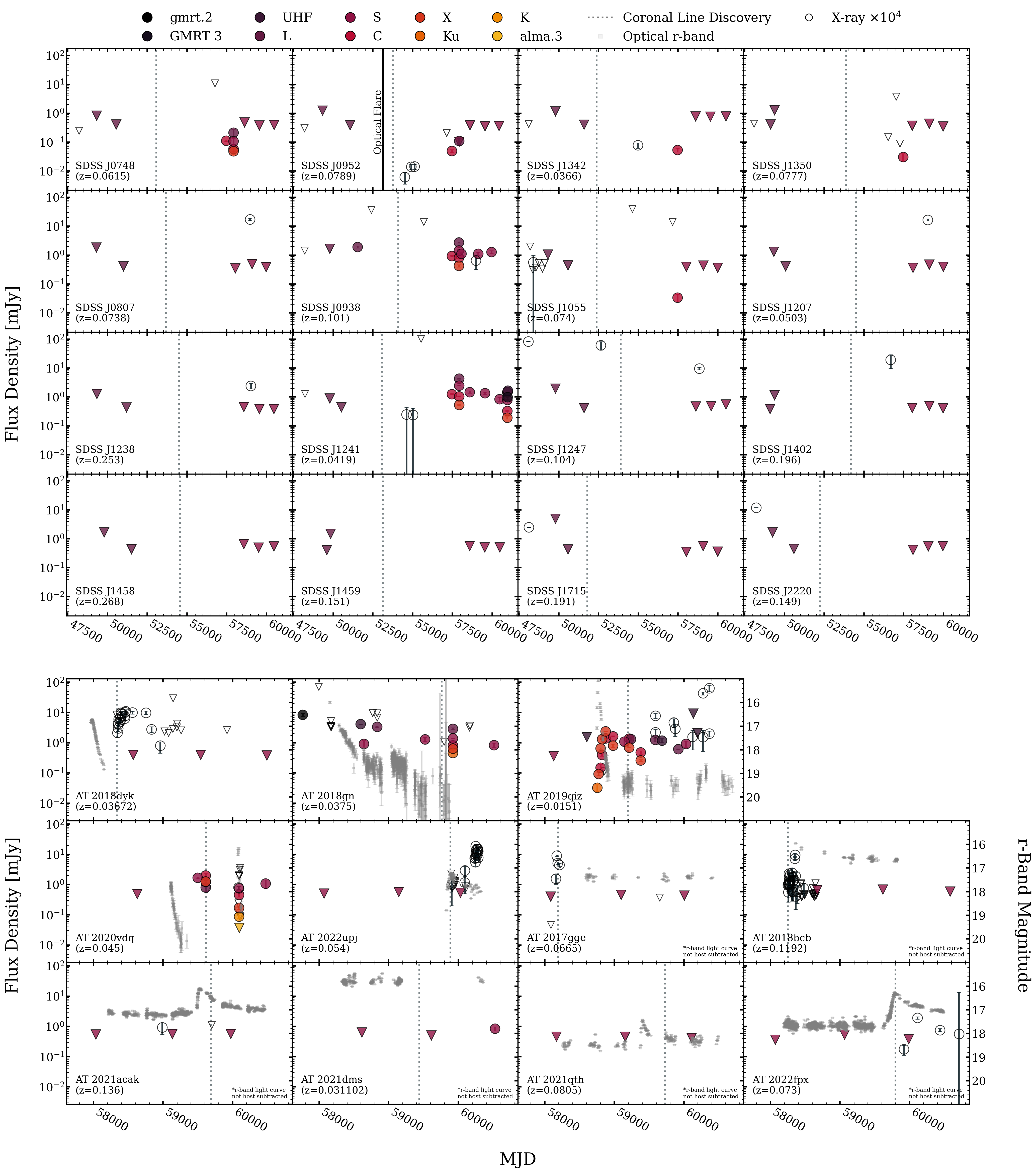}
    \caption{The multi-frequency light curves of the 27 ECLEs in our sample. The color of the point indicates the radio band of the observation and downward facing triangles indicate upperlimits resulting from non-detections in the continuum images. X-ray light curves are shown as open points. We show the optical r-band light curve for the 11 TDE-candidates as the right y-axis. The r-band light curves of AT\,2017gge, AT\,2018bcb, AT\,2021acak, AT\,2021dms, AT\,2021qth, and AT\,2022fpx are not host subtracted. For SDSS\,J0952 we also show a vertical line at the start of the optical flare discovered in \citet{palaversa_revealing_2016}. The name and redshift are labelled in the bottom left corner of each subpanel. For visualization purposes, we only show the NVSS and FIRST limits where constraining. Most ECLEs are not detected in the radio over multiple epochs. However, many of these upper limits are not constraining for the typical luminosity observed for the TDE population (see Figure \ref{fig:sband-lc}).}
    \label{fig:radio-lc}
\end{figure*}

In the following subsections we discuss both our targeted observations and archival searches for these 27 ECLEs. A summary of our targeted radio observations is in Table \ref{tab:obs}. The multi-frequency light curves for all 27 objects in flux space is in Figure \ref{fig:radio-lc}. The 3~GHz light curve in luminosity space for all 27 objects is in Figure \ref{fig:sband-lc}. The spectral energy distributions (SEDs) of ECLEs that have targeted multi-frequency observations are shown in Figure \ref{fig:sed-fits}. All reported limits are $3\sigma$. All data will be made publicly available on the Open mulTiwavelength Transient Event Repository \citep[OTTER;][]{ franz_python_2026,franz_open_2026}\footnote{\url{https://otter.idies.jhu.edu}}. The results of our targeted radio observations are reported in Appendix \ref{app:targeted-radio-obs}.

\begin{table}[]
    \centering
    \caption{Summary of Targeted Observations}
    \label{tab:obs}
    \begin{tabular}{lllll}
    \hline
         Target & Date & Tele- & Array & Band(s) \\
         & (UTC) & scope & Config. & \\
         \hline 
         SDSS\,J0748 & 2016 Mar 11 & VLA & C & S \\ 
         & 2017 Jun 10 & VLA & C & L,S,C,X \\ 
         \hline
         SDSS\,J0938 & 2016 Mar 12 & VLA & C & S \\ 
         & 2017 Jun 7 & VLA & C & L,S,C,X \\
         \hline
         SDSS\,J0952 & 2016 Mar 12 & VLA & C & S \\
         & 2017 Jun 18 & VLA & C & L,S,C,X \\
         \hline
         SDSS\,J1241 & 2016 Mar 12 & VLA & C & S \\
         & 2017 Jun 14 & VLA & C & L,S,C,X \\
         & 2025 Sep 26 & VLA & B & L,S,C,X \\
         & 2025 Oct 24 & GMRT & -- & 3,5 \\
         & 2025 Nov 7 & GMRT & -- & 4 \\
         \hline 
         SDSS\,J1342 & 2016 Mar 12 & VLA & C & S \\
         SDSS\,J1350 & 2016 Mar 12 & VLA & C & S \\
         SDSS\,J1055 & 2016 Mar 12  & VLA & C & S \\
         \hline
    \end{tabular}
\end{table}

\subsection{VLA Observations}
Radio observations of 7/16 ECLEs discovered in SDSS archival searches were taken with the VLA on March 11 and 12, 2016 (15B-247; P.I. Zauderer), roughly a decade post-ECL discovery. All seven ECLEs were observed with the the VLA C-band in the C configuration and detected with flux densities ranging from 30 to 1300 $\mu$Jy. Three of these ECLEs (SDSS\,J0748, SDSS\,J0938, SDSS\,J1241) were then followed up in June, 2017 (17A-368; P.I. Alexander) using the VLA bands L, S, C, and X. SDSS\,J0952 was also observed for a second epoch with the VLA in June, 2017 (17A-368; P.I. Alexander) using bands L and C. SDSS\,J1241 was followed up for a third epoch on 2025 September 26 (25B-242; P.I. Franz) using VLA bands L, S, C, and X. Data from all of these programs were reduced and imaged using standard procedures in pwkit \citep[][]{williams_variable_2017} and/or CASA \citep{mcmullin_casa_2007, casa_team_casa_2022}. All sources were detected at $>3\times$ the image RMS in all observations with the exception of the second epoch C-band observation of SDSS\,J0952, which resulted in an upper limit.

\subsection{GMRT Observations}

We also obtained an observation with the GMRT with Bands 3, 4, and 5 of SDSS\,J1241 (49\_043; P.I. Franz). The Band 3 and 5 observations were obtained on 24 October 2025 and the Band 4 observation was obtained on 07 November 2025, both quasi-simultaneous with our VLA observation. All data was reduced using standard procedures in CASA \citep[][]{mcmullin_casa_2007, casa_team_casa_2022}. We imaged each dataset with 3 rounds of phase only self-calibration and then fit a 2D gaussian the size of the beam to the source location to extract a flux density. The source was detected in all three observations.

\subsection{Publicly Available Radio Sky Surveys}
For all 27 ECLEs in our sample, we combine our targeted observations with observations from the VLASS quick look images \citep[][]{2020PASP..132c5001L,2023ApJS..267...37G}, the VLA FIRST \citep[FIRST;][]{1994ASPC...61..165B, 1995ApJ...450..559B} survey, and NRAO VLA Sky Survey \citep[NVSS;][]{condon_nrao_1998}.

The locations of all 27 objects has been observed with three epochs of the VLASS and with one epoch of the NVSS. $21/27$ events are in the FIRST footprint, with the 6 events outside of the footprint being AT\,2018gn, AT\,2019qiz, AT\,2022upj, AT\,2021dms, AT\,2018bcb, AT\,2021qth. 6/27 events are radio bright at some VLASS epoch and none of the events are detected in NVSS. The only FIRST image that has a detection is for SDSS\,J0938, which was initially reported in \citet{2011ApJ...737...45O}, and we opt to use their measured flux density (which is comparable within 1$\sigma$ to ours).

We extract a flux density from each image by fitting a 2D gaussian centered at the transient location and with a size fixed to the beam size. For the VLASS $ > 1.1$, NVSS, and FIRST images we use the integrated flux of the 2D gaussian. For any VLASS $= 1.1$ images, the CASA {\tt imfit} method fails to converge so we use the peak flux density of the 2D guassian. We consider a flux density a detection if it is $>5 \times$ the image RMS.

\subsection{Other Public Radio Data}
We query OTTER for other publicly available radio data for these events. OTTER includes a radio light curve of the TDE AT\,2019qiz from \citet{alexander_multiwavelength_2026}. OTTER also includes two SEDs and additional radio light curve points of the TDE AT\,2020vdq from \citet{somalwar_vlass_2025, somalwar_vlass_2025-1, somalwar_first_2025}. Finally, OTTER includes SEDs and a radio light curve for AT\,2018gn \citep{wang_asassn-18ap_2024}.

\subsection{Optical \lowercase{r}-Band Light Curves}
For comparison to our radio data we also include an optical r-band light curve for the 11 ECLEs discovered following a TDE candidate. First, we check OTTER \citep[][]{franz_open_2026} for a publicly available published r-band light curve. AT\,2018gn \citep[][]{wang_asassn-18ap_2024}, AT\,2018dyk \citep[][]{holoien_investigating_2022}, AT\,2019qiz \citep[][]{nicholl_outflow_2020}, AT\,2020vdq \citep[][]{mummery_fundamental_2024, somalwar_first_2025}, and AT\,2022upj \citep[][]{newsome_mapping_2024} are in OTTER and have host subtracted r-band light curves that we use. For the rest of the TDE-ECLEs (AT\,2017gge, AT\,2018bcb, AT\,2021acak, AT\,2021dms, AT\,2021qth, and AT\,2022fpx) we query the Zwicky Transient Facility (ZTF) archive using the OTTER Python API \citep[][]{franz_python_2026} and show the raw, not host-subtracted ZTF r-band light curve instead \citep[][]{masci_zwicky_2018, 2019PASP..131a8002B}.

\subsection{Archival X-ray Light Curves}
We first compile archival X-ray light curves from the literature. Querying OTTER for literature X-ray data provides light curves for SDSS\,J0748 \citep[][]{auchettl_new_2017}, SDSS\,J0952 \citep[][]{komossa_ntt_2009, auchettl_new_2017}, SDSS\,J1342 \citep[][]{auchettl_new_2017}, SDSS\,1350 \citep[][]{auchettl_new_2017}, AT\,2017gge \citep[][]{wang_discovery_2022}, AT\,2018gn \citep[][]{masterson_new_2024}, AT\,2018bcb \citep[][]{neustadt_tde_2020}, AT\,2018dyk \citep[][]{masterson_new_2024},  AT\,2020vdq \citep[][]{somalwar_first_2025}, AT\,2022fpx \citep[][]{lin_insights_2025}, and AT\,2022upj \citep[][]{newsome_mapping_2024}. We extract the AT\,2019qiz X-ray light curve directly from \citet{nicholl_quasi-periodic_2024} and an X-ray upperlimit on AT\,2021acak from \citet{li_at2021acak_2023}. For SDSS\,J0938, SDSS\,J1055, and SDSS\,J1241 we use the additional data presented in \citet{auchettl_new_2017}. The remaining X-ray data we use is from the European Space Agency's XMM-Newton Observatory Slew Survey \citep[XMM-Slew;][]{2008A&A...480..611S}, the Second ROSAT All-Sky Survey Point Source Catalog \citep[ROSAT;][]{2016A&A...588A.103B}, and the first release of the German eROSITA All-Sky Survey X-Ray Source Catalog \citep[eROSITA;][]{2024A&A...682A..34M}.

\section{Radio Spectral Energy Distribution Modeling} \label{sec:sed-model}
\begin{figure*}
    \centering
    \includegraphics[width=0.9\linewidth]{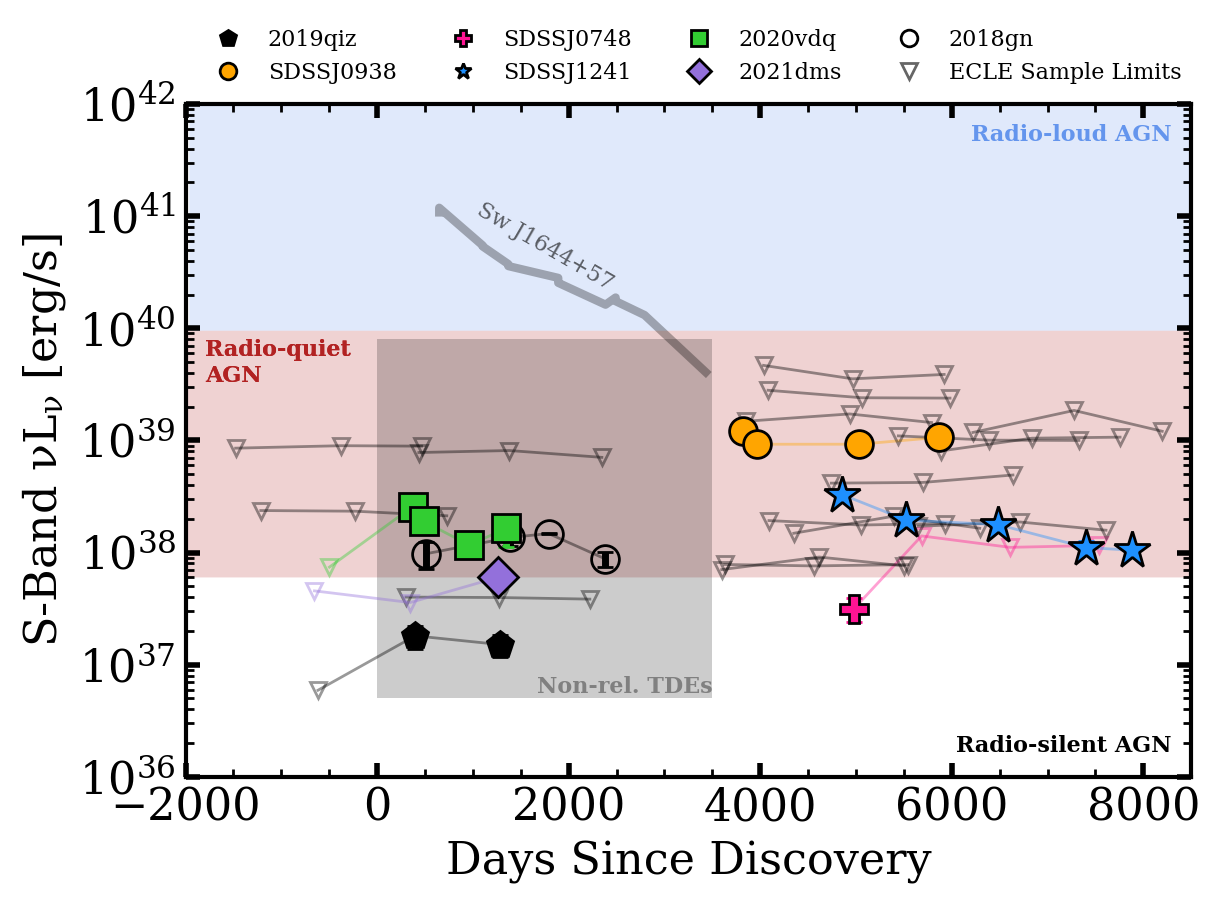}
    \caption{The S-band light curve in luminosity space of the ECLEs (colored) as compared to the broader TDE population from OTTER \citep[][]{franz_open_2026, franz_python_2026} and the AGN population \citep[][]{2016ApJ...831..168K}. The non-relativistic TDE sample (grey box) includes transients in OTTER with a classification confidence $C(TDE) > 0$ (i.e., events that are either photometrically selected or are spectroscopically confirmed), are optically discovered, and have at least one radio observation. The time is scaled to the discovery date of the transient. In the case of the ECLEs discovered in SDSS we use the coronal line discovery date as the transient discovery date \citep[][]{komossa_discovery_2008, wang_extreme_2012, yang_long-term_2013, clark_long-term_2024}. AT\,2018gn is shown as open circles because the radio emission is dominated by host star formation \citep{wang_asassn-18ap_2024} so they can be interpreted as upperlimits on the transient component of the radio emission. Qualitatively, the ECLE population has a similar S-band luminosity to both TDEs with non-relativistic outflows and radio-quiet AGN.}
    \label{fig:sband-lc}
\end{figure*}

\subsection{Basic Considerations}
Modeling the radio spectral energy distribution (SED) can inform our understanding of the emission mechanisms and physical properties of the system. Six of the ECLEs in our sample have at least one multi-band (i.e., observations with multiple radio receivers) epoch of observations with multiple detections: SDSS\,J0748 (1 epoch; this work), SDSS\,J0952 (1 epoch; this work), SDSS\,J0938 (1 epoch; this work), SDSS\,J1241 (2 epochs; this work), AT\,2018gn \citep[1 epoch;][]{wang_asassn-18ap_2024}, and AT\,2020vdq \citep[2 epochs;][]{somalwar_first_2025}. The radio emission detected from AT\,2018gn is resolved and associated with star formation \citep[][]{wang_asassn-18ap_2024}, so we do not model it here. SDSS\,J0952 also has a multi-frequency observation, but it only resulted in one detection, making spectral modeling impossible. Therefore, we are only able to model the multi-frequency observations of SDSS\,J0748, SDSS\,J0938, SDSS\,J1241, and AT\,2020vdq.

Based on the peaked nature of the SEDs (e.g., \autoref{fig:sed-fits}), the radio emission can be modeled as synchrotron radiation in one emitting region. Assuming that the radio outflow was launched around the discovery of the ECLs (2004-2006 for the SDSS ECLEs) or discovery of the TDE (in the case of AT\,2020vdq), we expect the emission to be dominated by synchrotron self-absorption without a significant cooled component, analogous to radio observations of TDEs (e.g., early-time observations \citealt{alexander_radio_2020}; late-time observations of Sw\,J1644+5, \citealt{cendes_radio_2021}; ASASSN-15oi, \citealt{hajela_eight_2025}; and other late-time brightening TDEs, \citealt{cendes_ubiquitous_2024}). 

Modeling of a synchrotron spectrum requires assumptions about the velocity and geometry of the outflow. 
We assume that the radio emission is a non-relativistic, spherically expanding outflow. This assumption is appropriate for both (1) non-relativistic outflows launched at any epoch and (2) promptly launched relativistic jets at any viewing angle at later epochs, as the emission is expected to have decelerated and transitioned to a quasi-spherical state $>10$ years after disruption \citep{nakar_detectable_2011, cendes_radio_2021}. For all four events that we model, the relatively low luminosity of the radio light curve (as compared to, e.g., Sw\,J1644+57) is consistent with a non-relativistic outflow expanding quasi-spherically into the CNM.

\subsection{Synchrotron Modeling}
To derive physical properties of the system, we first need to find the peak flux, frequency at peak, and optically thin power law decline of the SEDs. A synchrotron self-absorption peaked spectrum has an optically thick spectrum with $F_\nu\propto\nu^{5/2}$ and optically thin spectrum with $F_\nu\propto\nu^{(1-p)/2}$, where $p$ is the power-law index of the assumed power-law distribution of electrons. There is no evidence of additional breaks (e.g., cooling or minimal energy breaks) in the SEDs. This results in the spectral model described by equation \ref{eq:sed_model} \citep[e.g., as in ][]{granot_shape_2002}
\begin{align}
    F_\nu = F_{\rm \nu,p}\Biggl[ 
    \left(\frac{\nu}{\nu_{\rm a}}\right)^{-s\frac{5}{2}} + 
     \left(\frac{\nu}{\nu_{\rm a}}\right)^{-s\frac{(1-p)}{2}}\Biggr]^{-1/s} \;,\label{eq:sed_model}
\end{align}
\noindent where $F_{\rm \nu,p}$ is the flux of the synchrotron self-absorption break, $\nu_a$ is the frequency of the synchrotron self-absorption break, and $s$ is a smoothing parameter. At the self-absorption break, we assume the same smoothing parameters as past non-relativistic radio modeling papers of TDEs, $s = 1$ (e.g., \citealt{christy_dichotomy_2026})\footnote{This smoothing parameter differs from that prescribed in \citet{granot_shape_2002}: $s = 1.25-0.18p$. We choose to use $s=1$ because modeling the SEDs strictly following the \citet{granot_shape_2002} prescription resulted in unphysically large $p\approx4$ constraint for both multifrequency observations of SDSS\,J1241. This is not unexpected given that the smoothing parameters prescribed in \citet{granot_shape_2002} are derived from calibrating to ultra-relativistic jets, which is not the appropriate regime based on our assumptions.}.

\begin{figure*}
    \centering
    \includegraphics[width=\linewidth]{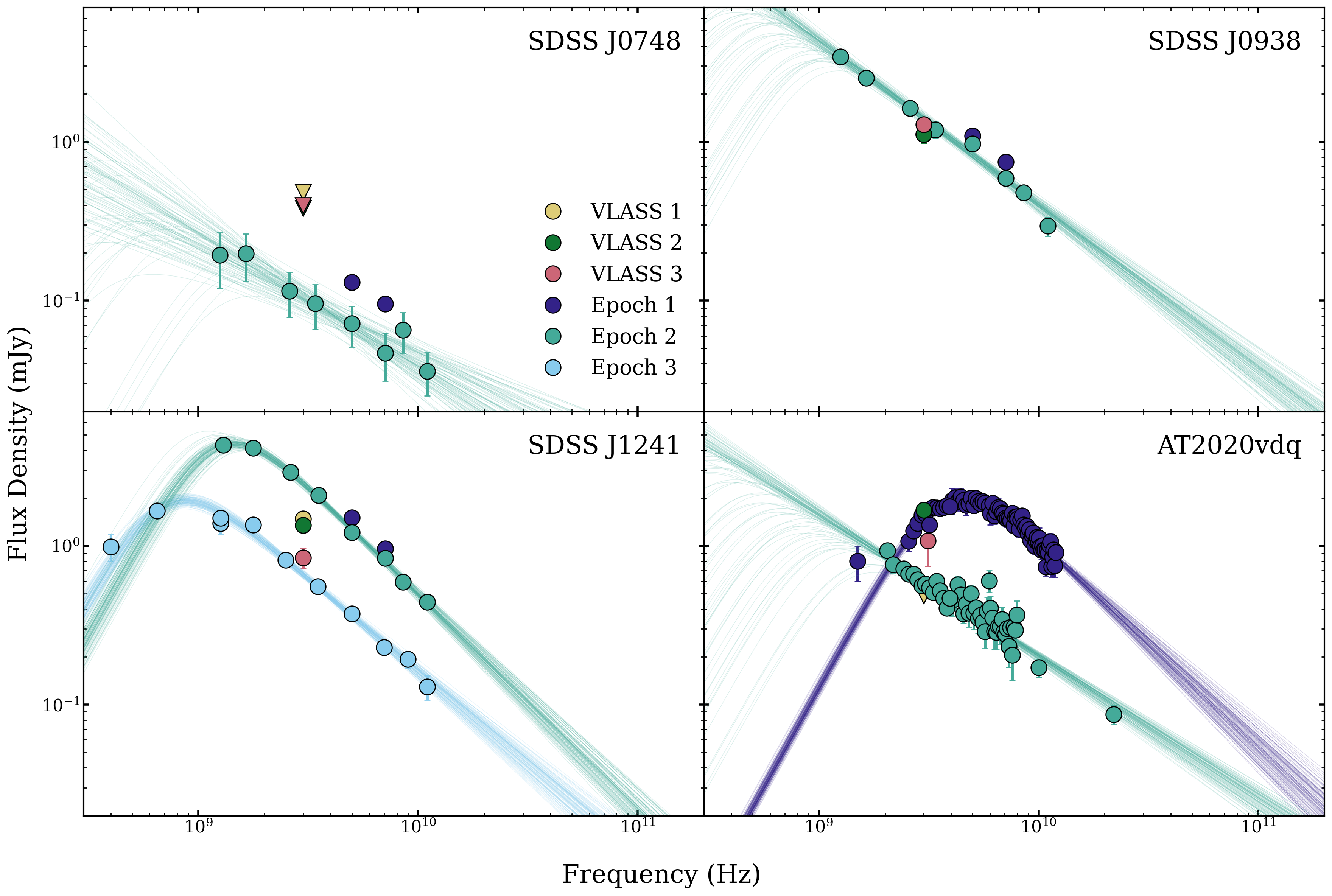}
    \caption{Four ECLEs in our sample have multi-band detections at the same epoch: SDSS\,J0748 (top left), SDSS\,J0938 (top right), SDSS\,J1241 (bottom left), AT\,2020vdq (bottom right). SEDs of all of the observations of these four events are shown here. The last 100 MCMC iterations are shown as a line on top of the data. All four events have multi-band epochs that are well fit as self-absorption peaked synchrotron spectra.}
    \label{fig:sed-fits}
\end{figure*}

To fit the four radio SEDs we use the \texttt{syncfit}\footnote{https://github.com/alexander-group/syncfit/tree/main} \citep[][]{franz_syncfit_zenodo} code with the {\tt B5} model, described by equation \ref{eq:sed_model}, and the \texttt{emcee} Markov-chain Monte Carlo backend fitting module \citep[][]{foreman-mackey_emcee_2013}. We leave $p$, $\nu_a$, and  $F_{\rm \nu,p}$ as free parameters with the uniform prior [2,4], log-uniform prior [6,11], and log-uniform prior [-4,2], respectively, where $F_{\rm \nu,p}$ is in mJy and $\nu_a$ is in Hz. For modeling we add a 5\% systematic uncertainty on the flux density in quadrature with the statistical uncertainty. We use a gaussian likelihood and treat upperlimits as hard limits on the prior (i.e., the model is not allowed to go above that point). All epochs are modeled with 100 walkers and 10,000 iterations, the first 1,000 of which are considered burn-in and not used during analysis. The broken power-law fits to the radio spectral energy distributions are in Figure \ref{fig:sed-fits}. 

\subsection{Modeling as a spherical non-relativistic outflow}

\defcitealias{duran_radius_2013}{BD13}

From the observationally derived parameters describing the SED, we assume equipartition following \citet{duran_radius_2013} (hereafter \citetalias{duran_radius_2013}) to derive the circumnuclear density, radius of the outflow, and energy of the outflow. Following \citet{cendes_radio_2021-1}, we add the appropriate correction factors for a non-relativistic (newtonian) outflow with $\Gamma \approx 1$. The equipartition radius is given in equation \ref{eq:radius} (\citetalias{duran_radius_2013} eq. 27),

\begin{widetext}
\begin{multline}
    R_{\rm eq} \approx 4^\frac{1}{(13+2p)} 10^{17}{\rm cm}~\left(21.8\right)^{\frac{1}{13+2p}} \left(525\right)^{\frac{p-1}{13+2p}} \gamma_m^{\frac{2-p}{13+2p}} \left[F_{\rm p, mJy}^\frac{p+6}{13+2p} d_{\rm L, 28}^{\frac{2(p+6)}{13+2p}} \nu_{\rm p, 10}^{-1} (1+z)^{-\frac{19+3p}{13+2p}}\right]
    f_A^{-\frac{5+p}{13+2p}} f_V^{-\frac{1}{13+2p}} \Gamma^{\frac{p+8}{13+2p}}  \;,\label{eq:radius} 
\end{multline}
\end{widetext}
where $p$ is the power-law index, $\gamma_m=2$ is the minimal electron energy lorentz factor, $F_{\rm p,mJy}$ is the peak flux in mJy, $d_{\rm L,28}$ is the luminosity distance in units of $10^{28}$ cm, $\nu_p$ is the frequency at the peak flux in Hz which we assume to be the self-absorption peak, $z$ is the redshift, $\Gamma = 1$ is the outflow lorentz factor for a newtonian outflow, $f_A=1$ is the area filling correction factor, and $f_V= \frac{4}{3} (1 - (1-f_{\rm emit})^3)$ is the volume filling correction factor. $f_{\rm emit}$ is the fraction of the outflow equipartition radius that is producing visible synchrotron emission. Following \citet{cendes_radio_2021-1} we assume $f_{\rm emit} = 0.1$, resulting in an $f_V \approx 0.36$. The extra pre-factor of $4$ appearing in equation \ref{eq:radius} and the following equipartition equations is a geometric correction for the Newtownian regime \citep[][]{duran_radius_2013, cendes_radio_2021-1}. The equipartition energy is given by equation \ref{eq:energy} (\citetalias{duran_radius_2013} eq. 28)

\begin{widetext}
\begin{multline}
    E_{\rm eq} \approx 4^\frac{11}{(13+2p)} \left(1.3\times10^{48}{\rm erg}\right)~\left(21.8\right)^{-\frac{2(p+1)}{13+2p}} \left(525\right)^{\frac{11(p-1)}{13+2p}} \gamma_m^{-\frac{11(p-2)}{13+2p}} \left[F_{\rm p, mJy}^\frac{14+3p}{13+2p} d_{\rm L, 28}^{\frac{2(3p+14)}{13+2p}} \nu_{\rm p, 10}^{-1} (1+z)^{-\frac{27+5p}{13+2p}}\right] \\ 
    f_A^{-\frac{3(p+1)}{13+2p}} f_V^{\frac{2(p+1)}{13+2p}} \Gamma^{-\frac{5p+16}{13+2p}}\;.\label{eq:energy}
\end{multline}
\end{widetext}

From equations \ref{eq:radius} and \ref{eq:energy}, the electron number density is given by equation \ref{eq:electrons} (modified from \citetalias{duran_radius_2013} eq. 15). $\gamma_e$ is the Lorentz factor of the electrons radiating at the peak frequency and is given by \citetalias{duran_radius_2013} eq. 14.

\begin{multline*}
    N_{e} \approx 4\times10^{54} \left(\frac{\gamma_e}{\gamma_m}\right)^{p-1} \\ \biggl[F_{\rm p,mJy}^3 d_{\rm L,28}^6 \nu_{\rm p,10}^{-5} \eta^\frac{10}{3} (1+z)^{-8} \biggr] f_A^{-2} R_{\rm eq, 17}^{-4}\;,
\end{multline*}
\begin{align}
    n_{e} \approx \frac{N_{e}}{V_{\rm emit}} \approx \frac{3N_{e}}{4\pi \left\{R_{\rm eq}^3 - \left[R_{\rm eq}\left(1 - f_{\rm emit}\right)\right]^3\right\}}\;, \nonumber \\ 
    n_{\rm ext} \approx \frac{n_e}{4\Gamma^2} \approx  \frac{3N_{e}}{16 \Gamma^2 \pi \left\{R_{\rm eq}^3 - \left[R_{\rm eq}\left(1 - f_{\rm emit}\right)\right]^3\right\}}\;.
    \label{eq:electrons}
\end{align}

\noindent Where $N_e$ is the number of synchrotron emitting electrons, $n_e$ is the number density of synchrotron emitting electrons, $n_{\rm ext}$ is the number density of synchrotron emitting electrons in the medium outside of the assumed strong shock, and $\eta = 1$ for the case where the peak frequency is the self-absorption frequency.

We then compute $R_{\rm eq}$, $E_{\rm eq}$, and $n_{\rm ext}$ for the entire posterior distributions of $p$, $F_\nu$, and $\nu_a$ from the broken power-law MCMC fits of the SEDs for SDSS\,J0748, SDSS\,J0938, SDSS\,J1241, and AT\,2020vdq. For well-constrained posteriors, we treat the resulting physical properties of the system as also well-constrained and use the median value of the distribution with uncertainties for the 16th and 84th quantiles. For poorly constrained parameters, we take the 95\% quantile as a limit. For self-consistency, we also recompute the equipartion values of literature TDEs based on the best fit SED parameters reported in those works, with the exception of Sw\,J1644+57 for which we directly use the values from \citet{eftekhari_late-time_2021}\footnote{This is necessary because the above equipartition equations do not directly apply for the Sw\,J1644+57 on-axis relativistic jet.}.

Finally, since we have two constraining observations of SDSS\,J1241, we solve \citetalias{duran_radius_2013} eq. 22 for $\beta$, the outflow velocity ($v$) in units of $c$, the speed of light.
\begin{align}
    \beta = \left[1 + \frac{c\Delta t}{\Delta R_{eq}(1+z)}\right]^{-1} \;.\label{eq:beta}
\end{align}
In equation \ref{eq:beta}, $\Delta t$ is the time between observations in the observer frame and $\Delta R_{eq}$ is the difference in the equipartition radius between the two observations. Applying this equation to SDSS\,J1241 yields a velocity $\beta \approx 0.006$ ($v \approx 1.8\times10^{3}~{\rm km/s}$). However, interpreting such a low velocity is non-trivial: If interpreted as the velocity of the outflowing material itself, we would then require an outflow mass of $\sim20~{\rm M_\odot}$ to match the kinetic energy implied by our calculated $E_{eq}$. This may be an artifact of our assumption of spherical symmetry, as using $f_A=0.1$ for a collimated outflow results in a more reasonable, although still large, outflow mass of a $\sim$few solar masses. Alternatively, this could be an indication that the radio emission is dominated by a compact region that lags behind the leading edge of the jet/outflow (analogous to the motion of ``hot spots" in resolved AGN jets, e.g., \citealt{mckean_lofar_2016}), allowing for the outflow to be moving faster than we infer from the radio emission. Overall, this low velocity likely indicates that our assumption of a spherical geometry in the case of SDSS\,J1241 is likely an oversimplification of reality. We nevertheless proceed with this assumption to facilitate direct comparison to the published TDE literature.

The median values, or $95\%$ quantile limits, for the model parameters are given in Table \ref{tab:results}. We show the electron number density profiles and outflow energy of the four ECLEs compared to the larger population of radio bright TDEs in Figure \ref{fig:n_e} and \ref{fig:E_outflow}, respectively. 

\begin{table*}
\centering
\caption{Equipartition Analysis Results}
\begin{tabular}{llllllll}
\toprule
Name & MJD & p & $\log_{10}\left(F_\nu/{\rm mJy}\right)$ & $\log_{10}\left(\nu_a / {\rm Hz}\right)$ & $n_e$ (cm$^{-3}$) & $R_{eq}$ (cm) & $E_{eq}$ (erg) \\ \\
\midrule
\multirow{2}{*}{SDSS\,J1241} & 60967.0 & $3.34^{+0.10}_{-0.10}$ & $0.557^{+0.021}_{-0.021}$ & $8.85^{+0.023}_{-0.024}$ & $14.5^{+3.51}_{-2.95}$ & $5.02^{+0.25}_{-0.23}\times10^{17}$ & $2.37^{+0.50}_{-0.41}\times10^{50}$ \\
 & 57959.0 & $3.72^{+0.11}_{-0.11}$ & $0.928^{+0.019}_{-0.019}$ & $9.09^{+0.027}_{-0.031}$ & $59.9^{+16.5}_{-14.1}$ & $4.53^{+0.34}_{-0.27}\times10^{17}$ & $7.26^{+1.29}_{-1.11}\times10^{50}$ \\
SDSSJ0938 & 57912.0 & $3.08^{+0.10}_{-0.10}$ & $>0.791$ & $<8.92$ & $<11.6$ & $>1.20\times 10^{18}$ & $>2.53\times 10^{51}$ \\
SDSSJ0748 & 57914.0 & $2.58^{+0.32}_{-0.30}$ & $>-0.556$ & $<9.00$ & $<20.8$ & $>1.40\times 10^{17}$ & $>5.22\times 10^{48}$ \\
\multirow{2}{*}{AT\,2020vdq} & 59612.0 & $3.60^{+0.17}_{-0.17}$ & $0.571^{+0.013}_{-0.014}$ & $9.59^{+0.016}_{-0.017}$ & $576^{+174}_{-140}$ & $1.04^{+0.02}_{-0.02}\times10^{17}$ & $8.41^{+2.69}_{-2.11}\times10^{49}$ \\
 & 60093.5 & $2.62^{+0.03}_{-0.03}$ & $>0.306$ & $<8.82$ & $<5.21$ & $>3.86\times 10^{17}$ & $>3.87\times 10^{49}$ \\
\bottomrule
\end{tabular}
\label{tab:results}
\end{table*}

\section{Results \& Discussion}\label{sec:discussion}

\subsection{Light curve properties}

The multi-wavelength light curve properties of our ECLE sample are detailed in the following subsections. The multi-wavelength light curves are shown in Figure \ref{fig:radio-lc} and we summarize our discussion of the light curves in Table \ref{tab:lc-summary}.

\begin{table*}[]
    \caption{Summary of Multi-wavelength Light Curve Properties}
    \centering
    \begin{tabular}{lp{0.7in}p{0.3in}p{0.35in}p{4.7in}}
        \hline
         & ECLE & Radio Det. & X-ray Det.\tablenotemark{\ddag} & Discussion Summary \& Other Remarks \\
         \hline
         \multirow{16}{*}[0cm]{\rotatebox{90}{\bf Non-variable ECLEs}} & SDSS\,J0938\tablenotemark{\dag} & \cmark & \cmark & The radio emission is non-transient. One X-ray detection combined with other non-constraining upperlimits makes interpretation of the X-ray emission difficult.  \\
         & SDSS\,J1715 & \xmark & \cmark & X-ray detected in one observation before the ECL appearance. Not detected in radio surveys. \\
        & SDSS\,J1055\tablenotemark{\dag} & \cmark & \cmark & Marginal X-ray detection. Faint radio detection from targeted follow-up. \\
        & SDSS\,J1459 & \xmark & --- & Not detected in radio surveys. \\
        & SDSS\,J0807 & \cmark & \cmark & Not detected in radio surveys with non-constraining limits. Detected in one epoch in the X-ray.\\
        & SDSS\,J1247 & \xmark & \cmark & Transient, decaying X-ray emission that was initially detected decades before the ECL discovery.\\
        & SDSS\,J1402 & \xmark & \cmark & X-ray detected in one observation after the ECL appearance. Not detected in radio surveys. \\
        & SDSS\,J1238 & \xmark & \cmark & X-ray detected in one observation after the ECL appearance. Not detected in radio surveys.  \\ 
        & SDSS\,J1207 & \xmark & \cmark & X-ray detected in one observation after the ECL appearance. Not detected in radio surveys. \\
        & SDSS\,J1458 & \xmark & --- & Not detected in radio surveys. \\ 
        \\ \hline \\ 
        \multirow{10}{*}[-3.8cm]{\rotatebox{90}{\bf Variable ECLEs}} & SDSS\,J2220 & \xmark & \cmark & X-ray detected in one observation after the ECL appearance. Not detected in radio surveys. \\
        & SDSS\,J1342\tablenotemark{\dag} & \cmark & \cmark & One epoch with a detection combined with non-constraining upperlimits in both the X-rays and radio.\\
        & SDSS\,J0748\tablenotemark{\dag} & \cmark & \xmark & The radio emission decays by a factor of $\approx2$ at C-band over two years.\\
        & SDSS\,J1241\tablenotemark{\dag} & \cmark & \cmark & Transient radio emission that increases by $\gtrsim10\times$ at 1.4~GHz between FIRST and our first targeted L-band observation. Two marginal X-ray detections with other non-constraining upperlimits. \\
        & SDSS\,J0952\tablenotemark{\dag} & \cmark & \cmark & It is unclear from our observations if the radio emission is transient. The X-ray emission appears transient \citep[][]{komossa_ntt_2009}.\\
        & SDSS\,J1350\tablenotemark{\dag} & \cmark & \xmark & One radio detection with other non-constraining upperlimits.\\
        & AT\,2017gge & \xmark & \cmark & Not detected in radio surveys. X-ray's appear slightly before the ECLs are discovered. \\
        & AT\,2018gn\tablenotemark{\dag} & \cmark & \xmark & Not detected in targeted X-ray follow-up. Detected in the radio but associated with host star formation \citep[][]{wang_asassn-18ap_2024}. ECLs appear within $\sim 4$ years after optical peak.\\
        & AT\,2018bcb & \xmark & \cmark & Not detected in radio surveys. ECLs likely appear sometime near optical peak. \\
        & AT\,2018dyk & \xmark & \cmark & Not detected in radio surveys. X-ray emission appears at the same time as the ECLs but after optical peak \citep[][]{masterson_new_2024}. \\
        & AT\,2019qiz\tablenotemark{\dag} & \cmark & \cmark & Transient radio emission that appears near optical peak and before ECLs. X-ray emission appears sometime after $\sim1$ year after optical peak \citep[][]{nicholl_quasi-periodic_2024}. \\
        & AT\,2020vdq\tablenotemark{\dag} & \cmark & \cmark & Two optical flares with ECLs appearing between them. Transient radio emission appearing within $\sim 2$ years of the first optical flare. X-ray detected after the second optical flare \citep{somalwar_first_2025}. \\
        & AT\,2021dms & \cmark & --- & Radio detected in VLASS with two preceding non-detections. \\
        & AT\,2021qth & \xmark & --- & Not detected in radio surveys. ECLs appear after the optical flare. \\
        & AT\,2021acak & \xmark & \cmark & Not detected in radio surveys. Detected in eROSITA {\it before} the ECL appearance. ECLs appear within $\sim 1$ year after optical peak.\\
        & AT\,2022fpx & \xmark & \cmark & ECLs appear near optical peak. Not detected in radio surveys. \\
        & AT\,2022upj & \xmark & \cmark & ECLs appear near optical peak. Not detected in radio surveys. X-ray emission is initially constant/non-detected but later brightens \citep[][]{newsome_mapping_2024}. \\
         \hline
    \end{tabular}
    \tablenotetext{\dag}{Detected with deeper targeted radio follow-up, which may bias the detection rate as compared to transients with observations only with radio surveys.}
    \tablenotetext{\ddag}{A ``---'' indicates no X-ray data available.}
    \label{tab:lc-summary}
\end{table*}

\subsubsection{X-ray}\label{sec:xray-lc}
There is limited X-ray temporal and spectral coverage of many of the ECLEs in this sample, with the exception of eight events: SDSS\,J0952, AT\,2018dyk, AT\,2017gge, AT\,2018bcb, AT\,2019qiz, AT\,2020vdq, AT\,2022fpx, and AT\,2022upj. All eight events are vECLEs and have time-variable X-ray emission, which seem to appear around the time of, or after, the ECLs are discovered\footnote{For SDSS\,J0952, the X-ray light curve shows brightening after ECL discovery, consistent with a brightening in the X-rays near the ECL discovery date. But, it does not have a constraining pre-ECL upper limit, making it unclear if SDSS\,J0952 also had pre-existing X-ray emission.}, well aligned with the expectation that a high flux of X-ray photons is necessary to produce such high-ionization gas \citep[e.g.,][]{komossa_discovery_2008}. Curiously, there appears to be a range of time delays between the appearance of the ECLs and the appearance of the X-ray emission. For example, the X-rays discovered following AT\,2018dyk seem to appear around the same time as the ECLs whereas the X-rays discovered following AT\,2022upj only seem to appear a $\sim$few hundred days after the ECLs. Some TDE models predict that, depending on the viewing angle, the initial X-ray emission may be obscured and re-radiated in the optical \citep[][]{guillochon_possible_2014,roth_x-ray_2016,dai_unified_2018, metzger_cooling_2022}. Such a model could explain a delay in the appearance of the X-ray emission without delaying the appearance of the ECLs \citep{thomsen_relativistic_2022}. However, it is unclear if the continuum of X-ray-ECL delay times is simply an observational artifact (i.e., AT\,2018dyk may not have received X-ray follow-up {\it until} the ECLs appeared). Future X-ray observations of ECLEs, both before and after the ECLs appear, are necessary to explore this potential continuum of X-ray-ECL delay times further.    
 
As an aside, we note that our search of the eROSITA point source catalog revealed an X-ray bright ($F_\nu \approx 8\times10^{-14}~{\rm erg~s^{-1}~cm^{-2}}$; $L_\nu \approx 4\times10^{42}$~erg/s) source potentially associated with the TDE candidate AT\,2021acak {\it before} the optical flare. This X-ray source is only 2.9'' from the optical coordinates, within the point spread function of the eROSITA survey. \citet{li_at2021acak_2023} reports an X-ray upper limit from Swift XRT $F_\nu \lesssim1.4\times10^{-13}~{\rm erg~cm^{-2}~s^{-1}}$, whereas this detection at $F_\nu \approx 8.2\times10^{-14}~{\rm erg~cm^{-2}~s^{-1}}$ is below this limit, indicating that the Swift observation was not sensitive to this low-level emission. Therefore, this is evidence of pre-existing nuclear activity in the host galaxy of AT\,2021acak, agreeing with \citet{li_at2021acak_2023} who interpret this event as a TDE in an AGN. Previous work has argued that evidence of pre-existing nuclear activity does not necessarily discount the TDE interpretation of the event (e.g., ASASSN-19bt, \citealt{christy_peculiar_2024} and PS16dtm, \citealt{blanchard_ps16dtm_2017}).

\subsubsection{Optical}
For the 11 post-TDE candidate ECLEs we have optical r-band light curves from all-sky surveys. The ECLs in these events appear anywhere from at peak optical light \citep[e.g., AT\,2022upj,][]{newsome_mapping_2024} to $\sim$years after peak optical light \citep[e.g., AT\,2019qiz,][]{short_delayed_2023}. AT\,2018bcb, AT\,2022fpx, and AT\,2022upj\footnote{and maybe also AT\,2017gge and AT\,2021dms, although ZTF does not capture the peaks of their light curves} all have variable ECLs that appear around the same time as peak optical light. However, all three of these events only have spectra beginning around peak light, making it unclear if the ECLs appeared prior to peak. In contrast, the variable ECLs in AT\,2018gn, AT\,2018dyk, AT\,2019qiz, and AT\,2021qth appear after the continuum optical light has faded beyond detectability or returned to the background host level. The vECLE AT\,2020acak appears to fall somewhere between these classes with the ECLs appearing during the decay of the optical light curve $\sim100$ days after peak. AT\,2020vdq is unique in that it has two flares, and the ECLs seem to appear between the two flares and remain, faintly, after the second flare \citep[][]{somalwar_vlass_2025-1, somalwar_first_2025}. This wide diversity in the ECL appearance time with respect to optical peak may indicate either a viewing angle effect, different 3D gas distributions (with rapid ECL appearance for significant amounts of gas along our line of sight and thus no light travel time delays with respect to the ionizing continuum, and longer light travel time delays otherwise), and/or an evolution in the optical--EUV--X-ray continuum shape over time (since it is the X-rays that produce the ions that emit the optical coronal lines). 

\subsubsection{Radio}
Many of the limits extracted from NVSS, FIRST, and VLASS are non-constraining for the typical non-relativistic radio emission observed from TDEs/ECLEs (\autoref{fig:sband-lc}). However, eight events have radio detections and/or constraining radio survey upperlimits, allowing us to discuss their nature in more detail: AT\,2018gn, AT\,2019qiz, AT\,2020vdq, AT\,2021dms, SDSS\,J0748, SDSS\,J0952, SDSS\,J1241, and SDSS\,J0938. The first seven of these events are vECLEs and, except for AT\,2018gn, have transient radio emission. In addition, AT\,2019qiz, AT\,2020vdq, AT\,2021dms, and SDSS\,J1241 have constraining pre-ECL detection radio upperlimits, indicating that the radio emission is likely powered by the same nuclear activity episode that triggered the ECLs. In contrast, the nvECLE SDSS\,J0938 has persistent 3~GHz radio emission. 

Of the seven vECLEs, AT\,2019qiz has promptly detected radio emission indicative of a non-relativistic outflow \citep[][]{nicholl_outflow_2020, anumarlapudi_radio_2024, alexander_multiwavelength_2026}. AT\,2021dms is detected in the radio, but only with VLASS. Similar to AT\,2019qiz, AT\,2020vdq and AT\,2021dms have constraining pre-TDE/ECLE VLASS upper limits, making it clear that the radio emission is variable and only appears after the discovery of the TDE. AT\,2021dms seems to only have bright radio emission $\sim500$ days after the appearance of the ECLs, but the radio light curve is poorly-sampled precluding a definitive statement (there is a constraining non-detection from VLASS right after the discovery of the ECLs from AT\,2021dms, and a lack of observations until the next VLASS observation where it is detected). AT\,2018gn has persistent, diffuse radio emission consistent with star formation \citep[][]{wang_asassn-18ap_2024}. Further investigation of the origin of the radio emission produced by these events is unfortunately not possible without more detailed, higher cadence radio observations. 

SDSS\,J0952, SDSS\,J0748, and SDSS\,J1241 are detected at late times, $\gtrsim 10$ years after the initial detection of ECLs. SDSS\,J0952 and SDSS\,J0748 do not have constraining upper limits from NVSS and FIRST, making it impossible to determine the epoch of the onset of the radio emission. Nevertheless, for SDSS\,J0748, we have multiple epochs that reveal dimming emission over a few years, likely pointing to a contribution from a transient source. Conversely, SDSS\,J1241 has a constraining upperlimit from FIRST, indicating that it has transient radio emission that appeared around the same time as the appearance of the ECLs.

\begin{figure}
    \centering
    \includegraphics[width=\linewidth]{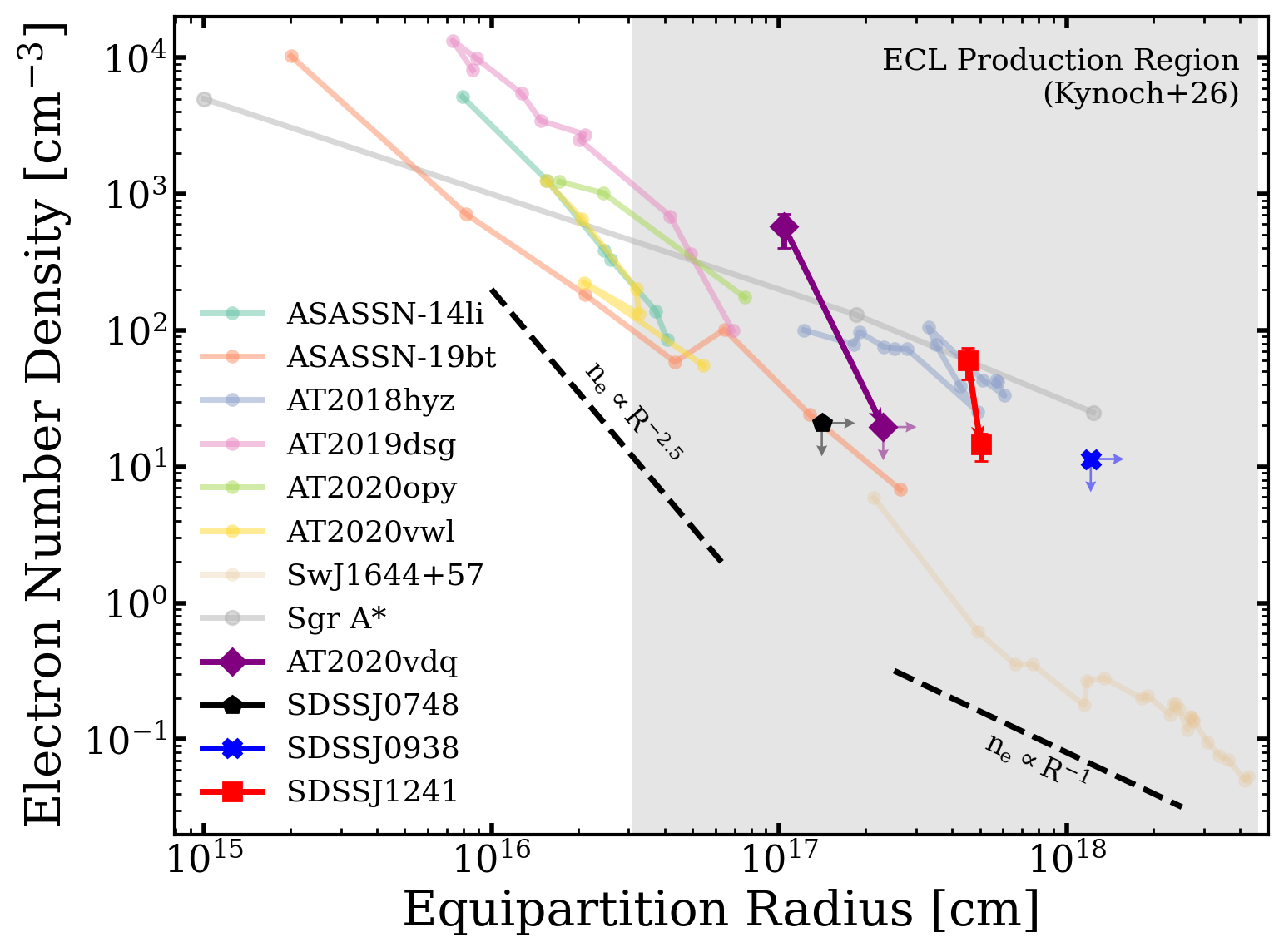}
    \caption{The electron number density profiles of the circumnuclear mediums for SDSS\,J1241, SDSS\,J0938\, and SDSS\,0748 as compared to the population of TDEs and Sgr A*  \citep{baganoff_chandra_2003, de_colle_dynamics_2012, alexander_discovery_2016,  gillessen_detection_2019, anderson_caltechnrao_2020, cendes_radio_2021, cendes_mildly_2022, eftekhari_radio_2018, short_delayed_2023, goodwin_radio_2023, goodwin_radio-emitting_2023, newsome_mapping_2024, cendes_continued_2026}. Limits are denoted as arrows pointing in the appropriate direction. The approximate radial distance from the SMBH where \citet{kynoch_mapping_2026} found the iron ECLs to emit from for SDSS\,J0748, SDSS\,J0938, and SDSS\,J1241 is shown as a grey shaded region. The three ECLEs shown on this figure have circumnuclear densities roughly consistent with the population of TDEs, M87*, and Sgr A*.}
    \label{fig:n_e}
\end{figure}

\begin{figure}
    \centering
    \includegraphics[width=\linewidth]{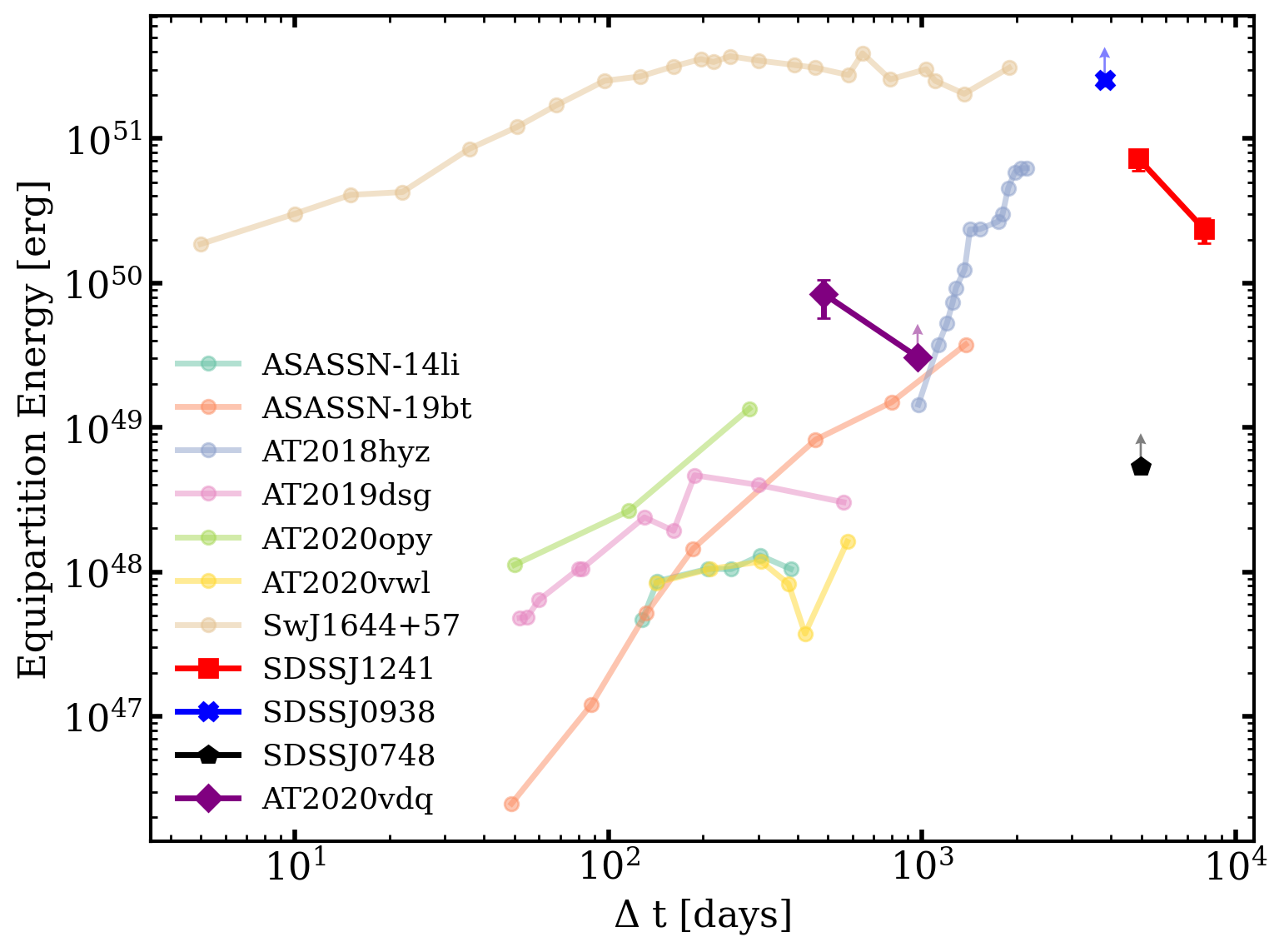}
    \caption{The equipartition energy as a function of days since discovery for SDSS\,J1241, SDSS\,J0938\, SDSS\,0748, and AT2020vdq as compared to the population of TDEs \citep{alexander_discovery_2016, anderson_caltechnrao_2020, cendes_radio_2021, cendes_mildly_2022, eftekhari_radio_2018, goodwin_radio_2023, goodwin_radio-emitting_2023, cendes_continued_2026}. Arrows on points indicate limits in that direction. For the three SDSS ECLEs the x-axis is the days since the discovery of the ECLs.}
    \label{fig:E_outflow}
\end{figure}

\subsection{The physical origin of ECLE radio emission}\label{sec:physical-origin}
We next attempt to explain the physical origin of the observed radio emission from the four best studied ECLEs in the radio: SDSS\,J0938, SDSS\,J0748, SDSS\,J1241, and AT\,2020vdq. We note that for all four events, our radio detections are much more luminous than typical radio emission from supernovae \citep[SNe, $\nu L_\nu \lesssim 10^{36}$~erg/s;][]{chomiuk_universal_2009}. Therefore, for all four ECLEs we are able to comfortably rule out a SNe origin for the radio emission. 

\subsubsection{nvECLE SDSS\,J0938: A persistent AGN jet}

Analyses of the optical spectra of SDSS\,J0938 in numerous past works reveal persistent AGN activity \citep[e.g.,][]{yang_long-term_2013, clark_long-term_2024}. In this work, we find that the radio emission is nearly constant over the past $\sim 20$ years and has a high outflow energy ($\gtrsim 10^{52}$ erg). Additionally, the detection of luminous ($L_\nu \approx 2\times10^{42}$~erg/s) X-rays in eROSITA likely points to the presence of an energetic accretion disk. These properties point to an AGN origin for the radio emission. 

\subsubsection{vECLE SDSS\,J0748: Star formation or a TDE?}
We have less information for SDSS\,J0748 as compared to SDSS\,J0938. We have 2 epochs of VLA C-band observations separated by $2$ years which reveal radio emission fading by a factor of $\approx 2$. The luminosity in the first epoch ($\nu L_{\nu = 1.4{\rm ~GHz}} \approx {\rm few}\times10^{37}~{\rm erg/s}$) is consistent with star formation \citep[][]{miller_radio_2009, kennicutt_star_2012, an_radio_2024}. Furthermore, the L-band luminosity indicates a reasonable SFR~$\approx 2.5 M_\odot/{\rm yr}$ \citep[][]{condon_radio_2002}. We compute the SFR from the narrow component of the H$\alpha$ flux found by \citet{wang_extreme_2012} and find a SFR~$\approx0.2 ~M_\odot/{\rm yr}$ \citep[][]{kennicutt_star_1998}. Given the $\sim$dex scatter in SFR relations, the possibility of optical extinction, and since radio emission probes star formation on longer timescale than H$\alpha$ \citep[][]{kennicutt_star_2012}, this radio emission is roughly consistent with star formation both in spectral index and luminosity. For completeness, we also check the MPA-JHU catalog \citep[][]{2004MNRAS.351.1151B} for SDSS\,J0748 and find a SFR~$=1.1^{+1.4}_{-0.6}~M_\odot/{\rm yr}$. This is also roughly consistent with the radio derived SFR.

However, if the radio was entirely due to star formation, we would expect the flux density to be constant. Instead, we see SDSS\,J0748 fade by a factor of $\sim2\times$ at $6$ GHz, indicating that a significant portion of the radio emission, in at least the first epoch, must have an origin besides star formation. We also see a lack of X-ray detections from this event in archival surveys. A fading of a factor of $\approx 2$ over 2 years in the radio and a lack of detectable X-ray emission is consistent with either an outflow produced by a TDE or low-level AGN variability \citep[e.g., ][]{hovatta_long-term_2008, alexander_radio_2020, cendes_ubiquitous_2024}.  

\subsubsection{vECLE SDSS\,J1241: Evidence for a powerful outflow}
SDSS\,J1241 has the most constraining information that a physical explanation of the radio emission must explain:
\begin{enumerate}
    \item Transient radio emission that is (a) not detected in 1.4 GHz observations from NVSS and FIRST prior to the discovery of the ECLs, (b) has decayed at 1-10 GHz over the past $\sim10$ years of radio monitoring (e.g., \autoref{fig:radio-lc}), and (c) A rise of a factor of $\gtrsim10$ between the FIRST upper limit and our brightest 1.4~GHz detection.
    \item An SED that (a) has a peak evolving to lower frequencies and lower flux densities over time and (b) is well fit by a self-absorbed peaked synchrotron spectrum.
    \item A slowly expanding outflow with high equipartition energy at late times, lying between the non-relativistic and relativistic TDE regimes (e.g., \autoref{fig:E_outflow}).
\end{enumerate}

The first two epochs of SDSS spectra from \citet[][]{wang_extreme_2012} and \citet{yang_long-term_2013} found no variation in the ECLs observed from SDSS\,J1241. However, \citet{clark_long-term_2024} followed up SDSS\,J1241 with Kast in 2021 and found no evidence of ECLs. Then, \citet[][]{kynoch_mapping_2026} followed up SDSS\,J1241 with DESI in 2025 and found evidence for faint, but still present, ECLs. These follow-up observations indicate that the ECLs are transient, but potentially on a longer timescale than other ECLEs. The WISE colors of SDSS\,J1241 are inconsistent with AGN and there are slight variations in the the long-term optical and IR light curves of SDSS\,J1241 \citep[][]{clark_long-term_2024}. However, an AGN origin can not be entirely ruled out, as emission line diagnostics (e.g., [\ion{O}{3}]/H$\beta$, [\ion{N}{2}]/H$\alpha$, and [\ion{S}{2}]/H$\alpha$) are consistent with AGN \citep[][]{clark_long-term_2024}. The difficulty in interpretation of the optical spectra of SDSS\,J1241 makes determining the origin of the radio emission more complex. 

The constraining NVSS ($F_\nu \lesssim 0.9$~mJy) and FIRST ($F_\nu \lesssim 0.4$~mJy) upperlimits at VLA L-band from 1995 and 1997, respectively, indicate that the radio emission turned on sometime between 1997 and 2017. This is a large time window, but this, combined with the continuously declining S-band radio emission since 2015, is convincing evidence that the radio emission observed from SDSS\,J1241 is transient and likely related to the discovery of the ECLs in the system on 2004 Feb. 27. The transient nature of the radio emission is in agreement with the transient nature of the ECLs observed from SDSS\,J1241 \citep[][]{clark_long-term_2024}. The initially persistent nature of the ECLs, but still transient on $\sim$decade timescales, in SDSS\,J1241 may indicate that the nuclear activity producing the ECLs was active on a longer time scale than in other ECLEs (e.g., AGN activity that we are watching ``turn off'' or a TDE with a long-lasting accretion disk).   

Given that the radio emission is transient and luminous ($\sim 10^{39}$ erg/s) on $\sim$decade timescales, we can comfortably rule out star formation as the source of the radio emission from SDSS\,J1241. Another possible explanation of the radio emission is an outflow with a delayed launch compared to the the ECLE appearance (i.e., an outflow launched just prior to 2015), as has been observed in some TDEs \citep[][]{cendes_ubiquitous_2024}. Such delayed outflows appear common in TDEs and are possibly explained as a state change in the accretion flow \citep[e.g., ][]{giannios_radio_2011, de_colle_dynamics_2012, goodwin_resolving_2026} or an off-axis jet decelerating into our line of sight \citep[e.g.,][]{sfaradi_off-axis_2024, cendes_continued_2026}. 

The high energy we derive reveals the presence of a powerful outflow. Since the radio properties of SDSS\,J1241 are most consistent with the late time ($\sim$2000 days since discovery) energies and luminosities of the on-axis jetted TDE Sw\,J1644+57 \citep[][]{cendes_radio_2021} and the likely off-axis jetted TDE\,2018hyz \citep[][]{cendes_mildly_2022,cendes_continued_2026}, SDSS\,J1241 may have harbored a relativistic jet in the past two decades. The low velocity of the outflow ($\beta\approx0.01$) complicates the interpretation of this event, but can be explained by a significant deceleration of a relativistic jet as it interacts with the ambient medium \citep{nakar_detectable_2011}. Or, such a low velocity combined with an energetic outflow may instead be evidence for a period of energy injection that suppresses the radio emission produced by an off-axis jet \citep{de_colle_beyond_2026}. The marginal detection of relatively faint X-rays at $L \approx2\times10^{41}$~erg/s may be in favor of the off-axis jet scenario \citep[e.g., as in TDE\,2018hyz; ][]{guolo_systematic_2024} since an on-axis jet would have luminous early-time X-ray emission that fades rapidly \citep[e.g., as in Sw\,J1644+57 which reached X-ray luminosities of $L \sim 10^{46}$~erg/s;][]{zauderer_birth_2011}, but the lack of X-ray coverage in the $\approx1000$ days after ECL discovery makes it impossible to test if the X-ray emission was persistently faint or initially very luminous.

Overall, there is evidence for a powerful outflow that originated from an on- or off-axis relativistic jet that has since undergone substantial deceleration and interaction with the surrounding medium. Thus, the most likely origin for the radio emission is an (on- or off-axis) jetted TDE or an AGN flare. The L-band increase of a factor of $\gtrsim10$ over $\sim 20$ years is an order of magnitude larger than a typical AGN flare \citep[$\sim0.1-2$; ][]{hovatta_long-term_2008}. We can therefore rule out normal AGN flaring and must instead invoke either a TDE or somewhat extreme AGN activity to explain the evolution of the observed radio luminosity and derived outflow energies. 

\subsubsection{vECLE AT\,2020\lowercase{vdq}: A TDE Driven Outflow}

AT\,2020vdq is widely accepted to be a TDE, and the observed radio emission is thought to originate from a Newtownian outflow produced by the TDE \citep[][]{somalwar_vlass_2025, somalwar_vlass_2025-1, somalwar_first_2025}. We agree with that interpretation of the radio emission here. We briefly note that it is curious that it has an equipartition energy (which we recompute self-consistently in this work) larger by a factor of a $\sim$few as compared to other non-relativistic TDEs at the same time. Although, it is the only ECLE where we have an energy constraint at relatively early times so we hesitate to draw any conclusions based on this observation.

\subsection{The fraction of radio bright ECLEs}\label{sec:fraction}

Based on the analysis in this work, we can place a rough lower limit on the fraction of all ECLEs that produce radio emission of $f \gtrsim 11/27 \approx 0.41$ (\autoref{fig:radio-lc}). Alternatively, if we instead only include vECLEs (see \autoref{tab:ecle-meta}), the fraction of vECLEs that produce radio emission is $f \gtrsim 9/17 \approx 0.53$\footnote{Or, if we include SDSS\,J1402, which is not radio detected and the nature of the observed coronal lines is subject to debate \citep[][]{callow_rate_2025}, this fraction becomes $f\gtrsim 9/18 \approx 0.50$.}. These are only lower limits on the fractions because many of the events in our sample have radio upper limits too shallow to rule out radio luminosities comparable to those of radio-detected TDEs and radio-quiet AGN (e.g., Figure \ref{fig:sband-lc}). 

Given the sensitivity of VLASS, a source must be brighter than $F_\nu \gtrsim 500\mu$Jy to be detected. We assume typical luminosities of: (1) non-relativistic radio outflows or radio-quiet AGN are $\nu L_\nu \approx 10^{38}$\,erg/s and (2) radio-loud AGN are $\nu L_\nu \approx 5\times10^{40}$\,erg/s. With VLASS we can detect radio-loud AGN at $z \lesssim 0.66$. All events in our sample are closer than this redshift limit, indicating that we can rule out a radio-loud AGN for the entire sample. With VLASS we can detect Newtownian outflows at $z \lesssim 0.04$. Only four events lie at $z \lesssim 0.04$: SDSS\,J1342, AT\,2019qiz, AT\,2021dms, and AT\,2018dyk. Only two of these are detected, indicating that the fraction of ECLEs with radio emission with $\nu L_\nu \gtrsim 10^{38}$\,erg/s to be roughly $\sim 50\%$. However, given the very limited sample size, we caution against making population level conjectures based on this result. 

For the 16 events with observations at $\delta t \gtrsim 1000$ days since discovery, we can also explore the presence of late-time luminous relativistic jets. The luminosity of an off-axis relativistic jet decelerating into our line of sight is $\nu L_\nu \sim 10^{39}$\,erg/s and with VLASS we can detect these events at $z \lesssim 0.12$. 10/16 of these events are closer than $z\approx0.12$. In section \ref{sec:physical-origin}, we find that the most likely explanation for the radio emission from two of these events (SDSS\,J0938 and SDSS\,J1241) is a relativistic jet, indicating that $2/10 \approx 20\%$ of archivally discovered ECLEs in our sample show evidence of a relativistic jet (or a comparably energetic outflow). However, we note that VLASS observations rule out a promptly launched relativistic jet for all 10 ECLEs discovered following TDEs (e.g., Figure \ref{fig:sband-lc}).

Finally, for ten events in our sample, we have deeper targeted VLA observations allowing us to detect radio emission brighter than $\approx30\mu$Jy, corresponding to $z \approx 0.15$ for typical Newtownian outflows. All ten events are closer than $z\approx0.15$ and we detect radio emission for all ten, indicating, based on a small sample size, that these ten ECLEs are radio bright with a $\nu L_\nu \gtrsim 10^{38}$ erg/s. However, once again, we strongly caution against extrapolating this fraction to the entire population given the small sample size. We also note that for 5/10 of these events, it is unclear whether the radio emission is driven by a persistent (e.g., star formation) or a transient (e.g., outflow or jet) source.

As a comparison to the fraction of radio bright ECLEs we consider three broader populations:
\begin{enumerate}
    \item Since transient ECLEs are thought to be associated with TDEs, we can compare to the broader TDE population. $\sim 40-60\%$ of TDEs exhibit radio emission at $<10$ years after discovery with a luminosity $\nu L_\nu \gtrsim {\rm few}\times10^{36}$ erg/s \citep[][]{alexander_radio_2020, cendes_ubiquitous_2024}.
    \item Since persistent ECLEs are though to be from AGN, we can compare to the optically-discovered AGN population. $\sim 80\%$ of AGN are radio-quiet, $\sim 20\%$ are radio-loud, and very few AGN are radio-silent \citep[][]{2016ApJ...831..168K}.
    \item Finally, we can also compare to the fraction of radio detected nuclear infrared flares, since many ECLEs also exhibit infrared flares \citep[][]{clark_long-term_2024, callow_rate_2025}. \citet{dai_compact_2020} find that $\sim 75\%$ of nuclear infrared flares in the mid-infrared outbursts in nearby galaxies \citep[MIRONG;][]{jiang_mid-infrared_2021} sample are radio detected.
\end{enumerate}
Based on the above discussion, we cautiously conclude that the fraction of radio bright transient ECLEs is $\gtrsim 50\%$. This constraint is entirely consistent with the fractions of AGN and nuclear infrared flares that are radio bright. The lower end of this constraint is roughly consistent with fractions of radio emission in the TDE population. However, observational biases (e.g., only four events are close enough to have constraining VLASS observations) limit the strength of these results. Additional, deeper, observations, particularly with better temporal coverage, of a larger sample of ECLEs are required to confirm this conclusion. 

The $\sim20\%$ incidence of relativistic jets discovered in this work for ECLEs may be better aligned with the fraction of radio-loud AGN ($\sim20\%$ of all AGN) rather than jetted TDEs ($\lesssim 1 \%$ of all TDEs). However, this may be due to an observational bias, as many of the events in our TDE comparison sample do not have radio observations at $\delta t \gtrsim 1000$ days after discovery, unlike our ECLE sample. As this is one of the first times that TDE candidates have been observed in the radio at at such late times, we may expect future late-time observations of TDEs to also reveal more off-axis jets than the rate previously reported in the literature. To test this prediction, we require multi-frequency radio observations of both TDEs and ECLEs on longer time scales with higher temporal coverage.

\subsection{Interpreting the radio-derived density}

Figure \ref{fig:n_e} shows the number density of synchrotron-emitting electrons as a function of radius for each of the ECLEs discussed in Section \ref{sec:physical-origin}. Our radio-derived density of the CNM ($n_e\sim100~{\rm cm^{-3}}$) is orders of magnitude smaller than the direct measurement of the density of the ECL-producing gas from spectral lines ($n_\ell\sim10^{6-9}~{\rm cm^{-3}}$, where $n_\ell$ indicates a density derived from spectral lines). If the radio-emitting region is assumed to be spherically symmetric, we find that the radio emission and the ECLs originate at comparable distances from the SMBH (shaded gray region in Figure \ref{fig:n_e}; \citealt{kynoch_mapping_2026}), indicating a disagreement in these two independent probes of the density.  Addressing this discrepancy is imperative for interpreting the radio-derived density.  

We first note that the radio emission and ECLs are not necessarily produced co-spatially. If the radio emission and the ECLs were produced co-spatially then both could originate from jet/outflow shock interaction with the CNM. However, the observed velocity widths of the optical ECLs are much too low for the ions to be produced by collisional ionization \citep{komossa_ntt_2009}. Therefore, the highly ionized metal ions must be produced through photoionization, while the radio emission is produced through shock interaction with the CNM. This indicates that the radio emission and ECLs provide complementary probes of the CNM and, when considered together, can provide insights into the geometry of the ECL-producing gas.

To address this discrepancy between $n_e$ and $n_\ell$, we first note that $n_e$ is the number density of relativistic electrons producing synchrotron emission and should not be taken as a direct measurement of the gas density. The vast majority of electrons present in the radio-emitting volume are not accelerated to relativistic energies by the shock and do not contribute appreciably to the observed radio emission. By making an assumption about the fraction of relativistic electrons $f_{\rm rel}$, we can convert $n_e$ to a total electron density, comparable to the gas density: $n_r \sim n_e/f_{\rm rel}$, where $n_r$ denotes a gas number density derived from radio observations. $f_{\rm rel}$ is poorly constrained by observations, but is generally agreed to be $f_{\rm rel} \ll 1$ \citep[][]{beniamini_constraints_2013}. 

If we assume $f_{\rm rel} \approx 0.01$\footnote{This is a conservatively chosen value based on constraints from gamma-ray bursts for $0.1 < f_{\rm rel} < 1$ \citep[][]{duncan_constraints_2023}. For a larger value (e.g., 10\%) the discrepancy between $n_r$ and $n_\ell$ would only be larger.}, this would imply $n_r \sim  10^4~{\rm cm^{-3}}$, which is still $\gtrsim2$ orders of magnitude smaller than the densities necessary for producing the ECLs. While calculating this radio-derived density, we assumed a spherical geometry, making the radio-derived density a measure of the ambient CNM rather than the ECL-producing regions themselves. Such a low radio-derived density thus indicates that the geometry of the CNM must be more complicated than our spherical model. 

To fully address the disagreement between the radio-derived and spectral-line-derived densities, the ECL-producing gas must be clumpy and may be distributed either in a dense torus or clouds of dense gas around the SMBH. In the toroidal case, the radio outflow would also probe a region of lower density gas perpendicular to the torus, whereas the ECLs are produced in the denser torus. Alternatively, a model with quasi-spherically distributed dense ECL-producing clouds also seems plausible. In this case, the shock powering the radio emission interacts with the same gas that is producing the ECLs, but our assumption of spherical symmetry is averaging out the densest regions. 

In either case, we can use our radio-derived density to place rough constraints on the volume filling factor ($f_V$) for the dense ECL-producing gas. Since we know the density of the ECL-producing region and that our radio-derived density is a spherical average probing the lower density gas, $f_V$ can be estimated as
\begin{align}
    f_{V} \sim \frac{n_r}{n_\ell} \sim \frac{n_{e}}{f_{\rm rel}\,n_\ell}\;.
\end{align} 
As mentioned above, we conservatively choose values of $f_{\rm rel}=0.01$ and $n_e=100~{\rm cm^{-3}}$. The lowest density regime constrained by spectral line measurements is $n_\ell \sim 10^{6}~{\rm cm^{-3}}$ \citep{komossa_discovery_2008, wang_extreme_2012}. Furthermore, \citet{komossa_ntt_2009} argued that it is reasonable to assume that the ECLs are emitted at densities not much above their critical densities, $n_\ell \lesssim 10^{9}~{\rm cm}^{-3}$, for the high-ionization species [\ion{Fe}{10}] and [\ion{Fe}{14}]. Therefore, using this range of possible $n_\ell$, we constrain $10^{-5}\lesssim f_{V}\lesssim10^{-2}$. 

Disentangling the toroidal and dense cloud geometric models requires a comprehensive analysis of multi-wavelength data. X-ray observations of AGN indicate that their torii are inhomogenous and have a fairly large covering fraction ($\sim 67\%$; \citealt{zhao_properties_2021}). This typical toroidal covering fraction is orders of magnitude larger than our radio-derived volume filling factor for the ECL-producing gas, indicating that we can confidently rule out a uniformly dense torus from producing the ECLs.\footnote{The only scenario in which a uniformly dense torus remains possible is if the radio emission originates in a very collimated jet that does not intersect the torus at all, rather than the $\sim$spherical thin shell that we assume. However, the implied density contrast between the torus and the ambient CNM would still be extremely large in this case.} Rather, the ECL-producing gas is likely concentrated in clumpy clouds that are located in either the torus or ambient, low-density surroundings. Future multi-wavelength observations of a much larger sample of ECLEs, combined with a rigorous statistical analysis, are necessary to fully understand the geometry and distribution of the ECL-producing gas.  

In summary, our radio observations exclude a dense spherical shield of ECL-producing gas surrounding the SMBH. Instead, these observations indicate that the ECL-producing gas is likely clumpy, residing in a much lower density CNM. Our finding of a low volume filling factor confirm that the assumption of $f_V \sim 10^{-4} - 10^{-5}$ used in \citet{mummery_galaxy_2025} is likely reliable. Future radio observations of a larger sample of ECLEs can provide improved independent constraints on $f_V$, better informing future modeling efforts of the ECL-producing gas.

\section{Conclusion}\label{sec:conclusion}

We present the first radio analysis of a population of 27 ECLEs. Four events (SDSS\,J1241, SDSS\,J0938, SDSS\,J0748, and AT\,2020vdq) have simultaneous multi-frequency detections allowing us to model their SEDs and extract physical properties of the system, including the energy of the outflow/jet and the density of the circumnuclear medium. We then compare the radio with the optical and X-ray light curve properties to determine the physical origin of the radio emission. We find a few notable observational characteristics of the ECLE population analyzed.
\begin{enumerate}
    \item $\sim50\%$ of ECLEs are detected in the radio. Even if we treat vECLEs separately this is still true. This fraction is roughly consistent with  the radio bright fractions of TDEs \citep[][]{alexander_radio_2020, cendes_ubiquitous_2024}, AGN \citep[][]{2016ApJ...831..168K}, and nuclear infrared flares \citep[][]{dai_compact_2020}.
    \item The radio emission from SDSS\,J0938 is likely from persistent AGN activity because of the high outflow energy and non-transient nature of its radio light curve. This is in good agreement with previous works.
    \item Due to the sparseness of the radio light curve, the origin of the radio emission from SDSS\,J0748 is more ambiguous, but we conclude that it is likely originating from low-level nuclear activity (either from a TDE or AGN).
    \item The late-time high luminosity ($\nu L_\nu \sim 10^{39}$ erg/s) and outflow kinetic energy ($\sim10^{51}$ erg) indicate that SDSS\,J1241 harbors a powerful outflow, potentially consistent with a transient relativistic jet that we are observing in a late-time non-relativistic phase. The consistency of the SDSS\,J1241 late time radio light curve and energy with the late-time evolution of the relativistic TDE Sw\,J1644+57 bolsters this finding. 
    \item The spherically averaged radio-derived circumnuclear electron density of SDSS\,J1241 is consistent with that of other TDEs in the ECL-producing region, and orders of magnitude smaller than the density derived directly from spectral line ratios. This indicates that the dense gas producing the ECLs is likely clumpy with a small volume filling factor ($10^{-5}\lesssim f_{V}\lesssim 10^{-2}$). 
\end{enumerate}

Confirming these results requires a larger sample size of radio observations of ECLEs with improved temporal coverage. In the case of TDE-ECLEs, these radio observations must first occur serendipitously as the ECLs seem to appear anywhere from at optical peak to years after optical peak. An alternative strategy is the radio follow-up of ECLEs discovered via archival searches of spectroscopic surveys \citep[e.g., DESI, ][]{clark_early_2026}; however, this limits the possibility of radio observations within $\sim$months after discovery, when the radio is probing the same region where the ECLs are produced. In the future, deep radio all-sky surveys with high temporal coverage, like the Deep Synoptic Array \citep[DSA;][]{hallinan_dsa-2000_2019}, will be necessary to follow-up large numbers of ECLEs and build a comprehensive sample.  

Broadly, radio observations of ECLEs provide a unique opportunity to probe the dense circumnuclear gas and the properties of outflowing material. Combining radio observations with X-ray and optical data directly connects the radio outflow and ambient circumnuclear gas to the ionizing continuum. In particular, the radio observations presented here provide a novel dynamical diagnostic that complements the photoionization signatures, revealing that the dense ECL-producing gas is clumpy and resides in a lower density CNM. As larger samples become available, ECLEs will become an essential laboratory for studying how sudden episodes of black hole accretion couples radiatively and mechanically to the gas within the central parsec of galaxies. 

\begin{acknowledgements}
    We thank Nathan Smith and Mathieu Renzo for useful discussions. N.F. acknowledges support from the National Science Foundation Graduate Research Fellowship Program under Grant No. DGE-2137419. K.D.A. is grateful for support from the Alfred P. Sloan Foundation. F.D.C. acknowledges support from the DGAPA/PAPIIT grant IN113424.

    The National Radio Astronomy Observatory and Green Bank Observatory are facilities of the U.S. National Science Foundation operated under cooperative agreement by Associated Universities, Inc. GMRT observations for this study were obtained via project 49\_043 (PI: Franz). We thank the staff of the GMRT who have made these observations possible. This work is based on observations carried out under project numbers E21AA and S22BS with the IRAM NOEMA Interferometer. IRAM is supported by INSU/CNRS (France), MPG (Germany), and IGN (Spain). This research has made use of data obtained from the XMM-Newton slew survey source catalogue compiled by the XMM-Newton Survey Science Centre consortium in collaboration with members of the XMM-Newton SOC and the EPIC consortium. This research has made use of data obtained from the Chandra Data Archive provided by the Chandra X-ray Center (CXC). We acknowledge the use of public data from the Swift data archive.
    
    This work is based on data from eROSITA, the soft X-ray instrument aboard SRG, a joint Russian-German science mission supported by the Russian Space Agency (Roskosmos), in the interests of the Russian Academy of Sciences represented by its Space Research Institute (IKI), and the Deutsches Zentrum für Luft- und Raumfahrt (DLR). The SRG spacecraft was built by Lavochkin Association (NPOL) and its subcontractors, and is operated by NPOL with support from the Max Planck Institute for Extraterrestrial Physics (MPE). The development and construction of the eROSITA X-ray instrument was led by MPE, with contributions from the Dr. Karl Remeis Observatory Bamberg \& ECAP (FAU Erlangen-Nuernberg), the University of Hamburg Observatory, the Leibniz Institute for Astrophysics Potsdam (AIP), and the Institute for Astronomy and Astrophysics of the University of Tübingen, with the support of DLR and the Max Planck Society. The Argelander Institute for Astronomy of the University of Bonn and the Ludwig Maximilians Universität Munich also participated in the science preparation for eROSITA. The UCSC team is supported in part by the Heising-Simons Foundation, the Vera Rubin Presidential Chair for Diversity at UCSC, and the National Science Foundation (AST-2307710, AST-2206243, AST-1911206).
    
     Based on observations obtained with the Samuel Oschin 48-inch Telescope at the Palomar Observatory as part of the Zwicky Transient Facility project. ZTF is supported by the National Science Foundation under Grant No. AST-1440341 and a collaboration including Caltech, IPAC, the Weizmann Institute for Science, the Oskar Klein Center at Stockholm University, the University of Maryland, the University of Washington, Deutsches Elektronen-Synchrotron and Humboldt University, Los Alamos National Laboratories, the TANGO Consortium of Taiwan, the University of Wisconsin at Milwaukee, and Lawrence Berkeley National Laboratories. Operations are conducted by COO, IPAC, and UW.

    Generate AI (GAI) was used for literature review \citep{perplexity_perplexity_2026} and improving visualizations \citep[e.g., choosing color schemes;][]{anthropic_claude_2026}. All GAI outputs were carefully verified through direct human supervision.  
\end{acknowledgements}

\facilities{
ADS,
VLA, 
GMRT,
CXO,
Swift,
ROSAT,
XMM,
eROSITA
}

\newcommand{\softwaredelim}{;~}
\software{
numpy \citep[]{harris_array_2020}\softwaredelim
matplotlib \citep[]{hunter_matplotlib_2007}\softwaredelim
astropy \citep{astropy_collaboration_astropy_2013, astropy_collaboration_astropy_2018, astropy_collaboration_astropy_2022}\softwaredelim
pandas \citep[]{reback2020pandas}\softwaredelim
emcee \citep[][]{foreman-mackey_emcee_2013}\softwaredelim
OTTER \citep[][]{franz_python_2026, franz_open_2026}\softwaredelim
synphot \citep{stsci_development_team_synphot_2018}\softwaredelim
pyarango\softwaredelim
astroquery \citep[]{ginsburg_astroquery_2019}\softwaredelim
pydantic \citep[]{Colvin_Pydantic_2025}\softwaredelim
fastapi \citep[]{Ramirez_FastAPI}\softwaredelim
CASA \citep[][]{mcmullin_casa_2007, casa_team_casa_2022}\softwaredelim
Claude \citep{anthropic_claude_2026}\softwaredelim
Perplexity \citep{perplexity_perplexity_2026}.
}

\appendix
\section{Targeted Radio Follow-up}\label{app:targeted-radio-obs}
The flux densities for all targeted observations that are presented in this work for the first time are in Table \ref{tab:targeted-radio-obs}. All photometry used in this work will be made publicly available on OTTER \citep[][]{franz_open_2026}.

\setlength{\extrarowheight}{0pt}
\renewcommand{\arraystretch}{0.9}

\begin{table}
\centering
\caption{Targeted Radio Observations} \label{tab:targeted-radio-obs}
\begin{tabular}{lrrlll}
\toprule
ECLE & MJD & $\nu$ (GHz) & Flux Density ($\mu$Jy) & Telescope & Project Code \\
\midrule
SDSS\,J0748 & 57458.92 & 5.00 & $130.30 \pm 13.50$ & VLA & 15B-247 \\
 & 57458.92 & 7.10 & $95.40 \pm 9.20$ & VLA & 15B-247 \\
 & 57914.72 & 1.25 & $193.40 \pm 73.90$ & VLA & 17A-368 \\
 & 57914.72 & 1.64 & $197.70 \pm 66.10$ & VLA & 17A-368 \\
 & 57914.72 & 2.60 & $114.80 \pm 36.60$ & VLA & 17A-368 \\
 & 57914.72 & 3.40 & $96.00 \pm 29.90$ & VLA & 17A-368 \\
 & 57914.72 & 5.00 & $71.60 \pm 20.60$ & VLA & 17A-368 \\
 & 57914.72 & 7.10 & $46.80 \pm 15.70$ & VLA & 17A-368 \\
 & 57914.72 & 8.55 & $65.40 \pm 18.70$ & VLA & 17A-368 \\
 & 57914.72 & 11.00 & $36.00 \pm 10.90$ & VLA & 17A-368 \\
\midrule
SDSS\,J0938 & 57459.04 & 5.00 & $1092.40 \pm 28.30$ & VLA & 15B-247 \\
 & 57459.04 & 7.10 & $745.30 \pm 25.50$ & VLA & 15B-247 \\
 & 57911.80 & 1.25 & $3421.80 \pm 196.30$ & VLA & 17A-368 \\
 & 57911.80 & 1.64 & $2526.60 \pm 130.90$ & VLA & 17A-368 \\
 & 57911.80 & 2.60 & $1629.10 \pm 159.70$ & VLA & 17A-368 \\
 & 57911.80 & 3.40 & $1191.90 \pm 141.70$ & VLA & 17A-368 \\
 & 57911.80 & 5.00 & $970.60 \pm 58.50$ & VLA & 17A-368 \\
 & 57911.80 & 7.10 & $589.70 \pm 37.90$ & VLA & 17A-368 \\
 & 57911.80 & 8.55 & $479.30 \pm 35.30$ & VLA & 17A-368 \\
 & 57911.80 & 11.00 & $295.40 \pm 40.40$ & VLA & 17A-368 \\
\midrule
SDSS\,J0952 & 57459.29 & 5.00 & $58.70 \pm 12.90$ & VLA & 15B-247 \\
 & 57459.29 & 7.10 & $34.70 \pm 8.50$ & VLA & 15B-247 \\
 & 57922.76 & 1.25 & $99.70 \pm 43.60$ & VLA & 17A-368 \\
 & 57922.76 & 1.64 & $101.20 \pm 33.50$ & VLA & 17A-368 \\
 & 57922.76 & 5.96 & $<105.09$ & VLA & 17A-368 \\
\midrule
SDSS\,J1055 & 57459.00 & 5.00 & $38.20 \pm 13.90$ & VLA & 15B-247 \\
 & 57459.00 & 7.10 & $30.50 \pm 12.20$ & VLA & 15B-247 \\
\midrule
SDSS\,J1241 & 57459.17 & 5.00 & $1502.00 \pm 20.10$ & VLA & 15B-247 \\
 & 57459.17 & 7.10 & $960.00 \pm 16.30$ & VLA & 15B-247 \\
 & 57918.05 & 1.30 & $4320.30 \pm 162.40$ & VLA & 17A-368 \\
 & 57918.05 & 1.77 & $4125.90 \pm 131.70$ & VLA & 17A-369 \\
 & 57918.05 & 2.63 & $2914.10 \pm 81.70$ & VLA & 17A-371 \\
 & 57918.05 & 3.52 & $2079.70 \pm 60.20$ & VLA & 17A-372 \\
 & 57918.05 & 5.00 & $1217.60 \pm 40.40$ & VLA & 17A-374 \\
 & 57918.05 & 7.10 & $837.50 \pm 31.10$ & VLA & 17A-375 \\
 & 57918.05 & 8.55 & $591.60 \pm 29.10$ & VLA & 17A-377 \\
 & 57918.05 & 11.00 & $443.30 \pm 37.20$ & VLA & 17A-378 \\
 & 60944.95 & 9.00 & $0.19 \pm 0.02$ & VLA & 25B-242 \\
 & 60944.95 & 11.00 & $0.13 \pm 0.02$ & VLA & 25B-242 \\
 & 60944.95 & 5.00 & $0.37 \pm 0.02$ & VLA & 25B-242 \\
 & 60944.95 & 7.00 & $0.23 \pm 0.02$ & VLA & 25B-242 \\
 & 60944.95 & 1.26 & $1.39 \pm 0.20$ & VLA & 25B-242 \\
 & 60944.95 & 1.78 & $1.36 \pm 0.06$ & VLA & 25B-242 \\
 & 60944.95 & 2.50 & $0.82 \pm 0.04$ & VLA & 25B-242 \\
 & 60944.95 & 3.50 & $0.55 \pm 0.03$ & VLA & 25B-242 \\
 & 60972.04 & 1.26 & $1.50 \pm 0.12$ & GMRT & 49\_043 \\
 & 60972.21 & 0.40 & $0.99 \pm 0.19$ & GMRT & 49\_043 \\
 & 60986.12 & 0.65 & $1.66 \pm 0.07$ & GMRT & 49\_043 \\
\midrule
SDSS\,J1342 & 57459.39 & 5.00 & $61.40 \pm 12.50$ & VLA & 15B-247 \\
 & 57459.39 & 7.10 & $45.20 \pm 12.00$ & VLA & 15B-247 \\
\midrule
SDSS\,J1350 & 57459.25 & 5.00 & $40.70 \pm 15.80$ & VLA & 15B-247 \\
 & 57459.25 & 7.10 & $23.40 \pm 9.20$ & VLA & 15B-247 \\
\bottomrule
\end{tabular}
\end{table}

\clearpage
\bibliography{main,otherrefs}{}
\bibliographystyle{aasjournalv7}

\end{document}